\documentclass{aa}  

\usepackage{graphicx}
\usepackage{txfonts}
\usepackage[separate-uncertainty=true,multi-part-units=single]{siunitx}
\usepackage{subcaption}
\usepackage[labelfont=bf]{caption}
\usepackage{amsmath}
\usepackage{placeins}
\usepackage{stfloats}

\begin{document}

   \title{Uncertainties of the 30-408 MHz Galactic emission \\ as a calibration source for radio detectors \\ in astroparticle physics}


   \author{M. Büsken \inst{1,2}
          \and T. Fodran \inst{3}
          \and T. Huege \inst{4,5}
          }

   \institute{Institute of Experimental Particle Physics, Karlsruhe Institute of Technology (KIT), Hermann-von-Helmholtz-Platz 1, 76344 Eggenstein-Leopoldshafen, Germany
   \and 
   Instituto de Tecnologías en Detección y Astropartículas, Universidad Nacional de San Martín, Av. General Paz 1555 (B1630KNA), San Martín, Buenos Aires, Argentina\\
    \email{max.buesken@kit.edu}
    \and
    Department of Astrophysics/IMAPP, Radboud University, P.O. Box 9010, 6500 GL Nijmegen, The Netherlands\\
    \email{t.fodran@science.ru.nl}
    \and
    Institute for Astroparticle Physics, Karlsruhe Institute of Technology (KIT), Hermann-von-Helmholtz-Platz 1, 76344 Eggenstein-Leopoldshafen, Germany
    \and
    Astrophysical Institute, Vrije Universiteit Brussel, Pleinlaan 2, 1050 Brussels, Belgium\\
    \email{tim.huege@kit.edu}
    }

   \date{Received / Accepted}
 
  \abstract
   {Arrays of radio antennas have proven to be successful in astroparticle physics with the observation of extensive air showers initiated by high-energy cosmic rays in the Earth's atmosphere. Accurate determination of the energy scale of the primary particles' energies requires an absolute calibration of the radio antennas for which, in recent years, the utilization of the Galactic emission as a reference source has emerged as a potential standard.}
   {To apply the "Galactic calibration," a proper estimation of the systematic uncertainties on the prediction of the Galactic emission from sky models is necessary, which we aim to quantify on a global level and for the specific cases of selected radio arrays. We further aim to determine the influence of additional natural radio sources on the Galactic calibration.}
   {We compared seven different sky models that predict the full-sky Galactic emission in the frequency range from \num{30} to \SI{408}{MHz}. We made an inventory of the reference maps on which they rely and used the output of the models to determine their global level of agreement. We subsequently took typical sky exposures and the frequency bands of selected radio arrays into account and repeated the comparison for each of them. Finally, we studied and discuss the relative influence of the quiet Sun, the ionosphere, and Jupiter.}
   {We find a systematic uncertainty of 14.3\% on the predicted power from the Galactic emission, which scales to approximately half of that value as the uncertainty on the determination of the energy of cosmic particles. When looking at the selected radio arrays, the uncertainty on the predicted power varies between 11.7\% and 21.5\%. The influence of the quiet Sun turns out to be insignificant at the lowest frequencies but increases to a relative contribution of $\sim30\%$ around \SI{400}{MHz}.}
   {}

   \keywords{Astroparticle physics --
            Methods: miscellaneous --
            Radio continuum: general --
            Sun: radio radiation
            }

    \titlerunning{Uncertainties of the 30-408 MHz Galactic Emission as a Calibration Source}
    \authorrunning{M. Büsken et al.}
    \maketitle

\section{Introduction} 
\label{sec:1_Introduction}
Already in the 1940s, surveys of the sky at radio frequencies were conducted. Strong emission from the constellation of Sagittarius was found as well as smaller maxima originating in extragalactic sources (e.g., Cygnus A and Cassiopeia A) and at this point also a signal coming from the Sun was seen \citep{Reber1944}. In measurement campaigns, the entire sky was mapped at specific frequencies. These maps show strong point-like sources as well as the diffuse radio emission from the Galaxy. The original motivation for this was to get a better understanding of the Galaxy and also to use the information from the maps in studies in the fields of astronomy and astrophysics, for example, related to the cosmic microwave background (CMB). To acquire accurate descriptions of the Galactic emission not only from a select number of maps at some specific frequencies but across a broad frequency range for the whole sky, efforts were made to build models of the radio sky based on these reference maps.

These sky models recently became relevant in the field of astroparticle physics for the absolute calibration of radio detection arrays. Over the past years, the radio detection technique has gained increasing importance for the observation of highly energetic cosmic rays and neutrinos detected through the radio emission from particle cascades \citep{Huege_2016, Schroeder_2017}. A number of promising radio detection arrays for the measurement of cosmic particles in the frequency range from a few tens to hundreds of megahertz are in the phase of development or under construction -- for example, the AugerPrime Radio Detector \citep{Pont_2019}, the Square Kilometer Array low-frequency site (SKA-low) \citep{Buitink_2021}, and the Giant Radio Array for Neutrino Detection (GRAND) \citep{GRAND_Design} -- while some are already taking data -- for instance, the Auger Engineering Radio Array (AERA) \citep{AERA}, and the LOw Frequency ARray (LOFAR) \citep{Schellart_2013}. For extracting physics it is very important to accurately determine the absolute energy scale of the detected particle cascades, and therefore an accurate calibration of the radio detectors is required. 

Calibrating with a reference antenna emitting a defined signal (e.g., mounted on a drone \citep{Aab_2017}) has the disadvantage of uncertainties on the emitted signal strength, which are difficult to assess \citep{Mulrey_2019}. Also, dedicated calibration campaigns require significant effort and are almost impossible to perform on a regular basis for large arrays. A different calibration approach uses the diffuse Galactic radio emission -- which poses a natural background to the detection of radio emission from particle showers -- as the reference signal. It typically poses the dominant background from natural sources at these frequencies \citep{ITU_Noise}. Although we often only refer to it as the Galactic emission in this work, we implicitly include subdominant extragalactic components as well. The calibration method based on the Galactic emission is called Galactic calibration and it also offers the opportunity to directly compare the calibrations of different radio arrays as they at least partially see the same sky. Moreover, the method can be applied as regularly as background data are available and it does not require dedicated field campaigns. The Galactic calibration was already applied to the LOFAR low-band antennas \citep{Mulrey_2019} and to the engineering array of the new radio detector of the Pierre Auger Observatory \citep{Fodran_2021}.

In the Galactic calibration, the measured background signal from the Galaxy is compared to predictions made with the aforementioned sky models. Knowing the uncertainty on these predictions is thus crucial for a useful absolute calibration. For this work, we therefore conducted a comparison of seven publicly available radio sky interpolation models for the frequency range from \num{30} to \SI{408}{MHz}. We did this by generating outputs with the models, calculating the average sky temperature from the outputs and determining the level of agreement between the models. From this comparison we got an estimate for the systematic uncertainties of the modeled background predictions for the Galactic calibration. 

First we present these models (Sect.\ \ref{sec:2_Models}) and summarize the reference maps on which they are based (Sect.\ \ref{sec:3_Maps}). Afterwards, we explain how we performed a global comparison while also showing other, less detailed descriptions of the radio background (Sect.\ \ref{sec:4_Comparison}). We subsequently adjusted the comparison for the sky seen by an observer at a specific location on Earth. We further conducted the comparison tailored to a set of selected radio arrays from the present and future, namely the Radio Neutrino Observatory Greenland (RNO-G) \citep{RNO-G_Design}, LOFAR \citep{Schellart_2013}, GRAND \citep{GRAND_Design}, Owens Valley Radio Observatory Long Wavelength Array (OVRO-LWA) \citep{Monroe_2020}, SKA-low \citep{Buitink_2021}, the radio detectors of the Pierre Auger Observatory \citep{AERA,Pont_2019}, and IceCube \citep{IceCube_SurfaceArray_Development}. Additionally, we studied the influence of other natural sources in the radio sky (Sect.\ \ref{sec:5_AdditionalSources}). Finally, we discuss implications of the results for the application of the Galactic calibration (Sect.\ \ref{sec:6_Discussion}) to radio detectors in astroparticle physics. 

\section{Radio sky interpolation models}\label{sec:SkyModels} 
\label{sec:2_Models}
In the past years, a couple of models for predicting the diffuse foreground emission of the sky in the radio and microwave regimes were developed with the purpose of calibrating radio arrays and to be used in foreground removal for 21-cm cosmology \citep{Hibbard_2020, Anstey_2021}. Motivation for developing these models was also given through efforts to map the CMB with high accuracy \citep{Tegmark_2000}. Moreover, radio sky models can be used for better estimating fluxes from pulsars and fast radio bursts \citep{Price_2021}.

These models interpolate between reference sky surveys at various frequencies conducted with telescopes at different locations. In the surveys, the sky brightness at given coordinates is mapped in terms of the brightness temperature $T_{\text{B}}$. This is the temperature of a thermal radiator, that is, a black body, that would show the same brightness as the one measured. In the classical limit $h\nu \ll k_{\text{B}}T_{\text{B}}$ of Planck's law for black body radiation, where $h$ is the Planck constant, $\nu$ is the considered frequency and $k_\text{B}$ is the Boltzmann constant, the Rayleigh-Jeans law is applicable. The brightness temperature is then directly proportional to the observed brightness $I_{\nu}\:[\mathrm{W m}^{-2} \mathrm{Hz}^{-1} \mathrm{sr}^{-1}]$:\citep{RadioAstroTools}

\begin{equation}
	T_{\text{B}} = \frac{c^2}{2k_{\text{B}} \nu^2}I_{\nu},
\end{equation}

where $c$ is the speed of light in vacuum. In the frequency range from a few $\SI{10}{MHz}$ to a few $\SI{100}{MHz}$, most of the electromagnetic background in the sky is presumably synchrotron radiation from electrons gyrating in the magnetic field of the Galaxy \citep{Rybicki_1979}. Although this is nonthermal emission, the description of the radio sky by a brightness temperature is still practical because of its proportionality to the brightness $I_{\nu}$. The frequency dependence of the brightness temperature can be described at lower radio frequencies by a power-law

\begin{equation}\label{eq:power_law}
	T_{\text{B}} \propto \nu^{\beta}
\end{equation}

with a spectral index $\beta$. Recent studies of the spectral index at frequencies from $\SI{50}{MHz}$ to $\SI{200}{MHz}$ lie in the range $\num{-2.62} < \beta < \num{-2.46}$, depending on the region in the sky \citep{Mozdzen_2016,Mozdzen_2019,Rogers_2008}. Around $\SI{200}{MHz}$ and again above $\SI{400}{MHz}$, changes of the spectral index are observed with a steepening towards higher frequencies \citep{Purton_1966, Bridle_1967, Webster_1974}.

This power-law relation can be used to scale a single full-sky reference map from a survey conducted at a low frequency to any other frequency. One model, called LFmap, uses this approach of spectral scaling with frequency- and region-specific spectral indices for interpolation. LFmap is included in the model comparison of this work. Other popular interpolation models use the approach of a principal component analysis (PCA) of multiple reference maps to produce frequency-dependent descriptions of the radio sky. In this inventory, three such models are considered, namely the Global Sky Model (GSM) in its first \citep{de_Oliveira_Costa_2008} and improved \citep{Zheng_2016} version, as well as the Low Frequency Sky Model (LFSM) \citep{Dowell_2017}. Lastly, three more models are included in this study which assume a physical model of the diffuse radio emission, that they fit to reference maps. These are the Global MOdel for the radio Sky Spectrum (GMOSS) \citep{GMOSS}, the high-resolution Self-consistent whole Sky foreground Model (SSM) \citep{SSM}, and the UltraLong-wavelength Sky Model with Absorption (ULSA) \citep{ULSA}. A short summary of all models, their core modeling approach, and whether they include the CMB contribution is given in Table \ref{tab:sky_models}.

\subsection{LFmap} \label{sec:LFmap}
The LFmap software by \citet{LFmap} is a radio sky interpolation software based on the simple power-law model for the sky brightness temperature (Eq.\ \ref{eq:power_law}). The \SI{408}{MHz} map by \citet{Haslam_1982} in the revisited version by \citet{Platania_2003} is scaled down with spectral indices from a spectral index map from the same reference to frequencies down to a minimum frequency $f_\text{bend}$. To take spectral bending into account, new spectral indices are used for frequencies below the bending frequency $f_\text{bend}$, which are calculated such that each pixel of the downscaled map at $f_\text{bend}$ scales onto the same pixel of a \SI{22}{MHz} map by \citet{Roger_1999}. Part of the region around the Galactic center is affected by HII absorption, which becomes relevant only below \SI{45}{MHz}. Therefore, an intermediate step at the absorption frequency $f_\text{absorption}$ is introduced, where the affected region of the map is replaced by the average of the surrounding area, that is unaffected by HII absorption. Below $f_\text{absorption}$ the procedure of calculating spectral indices is done with the original \SI{22}{MHz} map. In a similar manner, regions not covered in the \SI{22}{MHz} map are replaced with averages from surrounding areas. Most point-like sources are not treated individually, except for the two brightest ones, Cassiopeia A and Cygnus A, for which adapted spectral indices are used to scale their brightness temperatures. In this study, we used the default settings of LFmap, or more specifically, $f_\text{bend} = \SI{180}{MHz}$ and $f_\text{absorption} = \SI{45}{MHz}$.

\subsection{GSM (2008)} \label{sec:GSM}
Three other considered radio sky interpolation models use a principal component analysis (PCA) to generate all-sky maps at any intermediate frequency. Of these models, GSM \citep{de_Oliveira_Costa_2008} is the oldest, which uses the fewest reference maps. The advantage of a PCA is that also maps of limited sky coverage can be used rather easily. However, different angular resolutions and levels of quality between the used maps pose challenges for this approach. In GSM, also maps at gigahertz frequencies are used, which can have a significant influence on the low-frequency results in regions where there is otherwise only sparse map coverage. The sky temperature component from the CMB was subtracted from the reference maps before building the model. In our analysis, we added this component again to the generated maps.

The PCA in GSM is used to decompose the spatial structures of the diffuse radio sky from the information of the 11 input maps. The algorithm was performed for the region of the sky that is covered by all of the reference maps. In a fit, the principal components were modeled to best match in the remaining sky regions. At minimum, there is information from six maps per pixel. However, for a large part of the southern hemisphere, sub-gigahertz information is only available from the all-sky maps at \SI{45}{MHz} and \SI{408}{MHz}, possibly reducing the model accuracy in this region at these frequencies, which is relevant for experiments located in the southern hemisphere. The authors used three principle components for the final model, because they are sufficient to explain almost all of the emission seen in the reference maps. These first three principal components were also interpreted as to how they relate to the physical components of the diffuse radio emission. 
In order to be able to generate full sky maps with these principal components at any frequency between \SI{10}{MHz} and \SI{100}{GHz}, the components were fitted as a function of the logarithmic frequency using a cubic spline. This way, frequency spectra for each of the principal components were obtained. No special treatment for point-like sources is mentioned although they were removed in some of the used reference maps.

The accuracy of the model was determined by taking one of the reference maps out of the calculation of the principal components and calculating the difference of this map to the output of the model at that frequency without the information from the map. This was done for all frequencies at which a reference map is available leading to an estimate for the accuracy of around \SI{12}{\%} for the sub-gigahertz maps. However, the model does not account for individual uncertainties associated with the reference maps or systematic temperature shifts of a map over the whole sky, that could be introduced by inaccurate calibration.

\subsection{GSM (2016)} \label{sec:GSM16}
The original GSM received an update in which further reference maps were included and the model algorithm around the PCA was improved \citep{Zheng_2016}. The frequency range of the model was extended to \SI{5}{THz} by including recently published surveys in the microwave regime. At lower radio frequencies, maps at \SI{85}{MHz} and \SI{150}{MHz} were added. With this enhanced set of surveys, there is no common region covered by all of them, that could be used for conducting the PCA as in the original GSM. Instead, an iterative algorithm was implemented, which fits the principal components to the reference maps to the desired accuracy. For the zeroth iteration, a subset of maps was used that overlap by at least \SI{5}{\%} in sky coverage. Then, the remaining maps were fed into the model. Finally, six principal components were used, compared to three components in the original GSM. Within the PCA algorithm, point-like sources are removed.

A level of model accuracy is determined in the same way as for the original GSM and found to be around \SI{8}{\%} for the sub-gigahertz frequencies. When compared to the original GSM, the accuracy of the improved model is better for all sub-gigahertz frequencies by a factor of up to 2. However, since the original GSM is still widely used and since most of the newly added reference maps are at higher frequencies than we considered here, we included both versions of the GSM in the comparison of this study.

\subsection{LFSM} \label{sec:LFSM}
Based on the sky surveys at the Long Wavelength Array Station 1 (LWA1) \citep{Ellingson_2013} in the USA at frequencies from \SI{40}{MHz} to \SI{80}{MHz}, a low-frequency sky model was constructed \citep{Dowell_2017}. Additional surveys from the literature were included. The model itself is based on the PCA approach as performed in the original GSM. Similar to the improved GSM, an iterative procedure was used to tackle the problem of not having a sky region that is commonly covered by all maps. The final number of principal components in the model is three and maps can be generated within a total range from $10$ to \SI{408}{MHz}. Same as for GSM, no individual treatment of point-like sources is mentioned.

As a measure of accuracy, the relative difference between the LWA1 survey at \SI{74}{MHz} and the model prediction at the same frequency was calculated. Deviations are found to be at an overall level of \SI{10}{\%}. The LWA1 surveys nicely cover the low-radio regime below \mbox{$\SI{100}{MHz}$}. However, the LWA1 is stationed in the northern hemisphere. This could lead to additional uncertainties in the interpolated maps at Galactic latitudes below \SI{-40}{\degr}, which would become relevant for experiments in the southern hemisphere. Furthermore, the temperature scale of the LFSM output strongly depends on the calibration of the LWA1 surveys, as these represent a large fraction of the total ensemble of input maps. Potential systematic errors in the calibration would thus directly bias the interpolation model.

\subsection{GMOSS} \label{sec:GMOSS}
GMOSS is based on a physical model of the mechanisms of the diffuse radio emission in the sky, most importantly synchrotron radiation, thermal radiation, and free-free emission. This model is the basis for a parameterization of the sky brightness temperature spectrum. The parameterization was fit to the data from six reference maps for each pixel of a full-sky map at \SI{5}{\degree} resolution. The CMB contribution to the sky temperature was subtracted from the reference maps, which we added again to the generated maps in our comparison. Four of the reference maps are measured ones from \SI{150}{MHz} to \SI{23}{GHz}, while the other two were generated with GSM, at \SI{22}{} and \SI{45}{MHz}. These generated maps are assumed to be close to the reference maps at the same frequencies, that were used to build GSM but are not full-sky. However, this way a level of correlation between the outputs of GSM and GMOSS is introduced. The fit parameters of the GMOSS model were constrained to stay within physically reasonable limits. The fractional deviation between the fitted model and the reference data is within \SI{17}{\%} for most of the pixels.

\subsection{SSM} \label{sec:SSM}
The SSM is a model of the radio sky that separately estimates the diffuse emission and the emission from point-like source and adds them up. The temperature of the diffuse emission is modeled with the usual spectral power law, while the spectral index is taken as frequency- and direction-dependent. This dependency was accounted for by expanding the spectral index first in the logarithm of the frequency and then setting up the resulting relation to the sky temperature for each pixel in the sky. The parameters of this polynomial were then fitted using data from in total 13 reference maps, where the \SI{408}{MHz} map from \citet{Remazeilles_2015} acted as the basis map. The \SI{408}{MHz} map was scaled to any frequency using the power law with the fitted form of the spectral index. 

Prior to the fit, point sources were removed from those reference maps which originally included them. In SSM, point sources are separately modeled onto the diffuse emission with two catalogs. The spatial distribution of the point sources, as well as the spectral behavior of their emission were determined from these catalogs, which are at two different frequencies, \SI{843}{MHz} and \SI{1.4}{GHz}. Finally the diffuse emission and the emission from the point sources were added up for the final model output.

\subsection{ULSA} \label{sec:ULSA}
Special emphasis on ultra-long wavelength radio emission down to frequencies around \SI{1}{MHz} is given in the work of the ULSA model. In ULSA, free-free absorption in the Galaxy is taken into account that affects the brightness spectrum at frequencies below \SI{10}{MHz}. Nevertheless, the model can also be used to generate maps up to \SI{408}{MHz}.

In the model, the sky temperature is assumed as a sum of a Galactic and an isotropic extragalactic component with a reduction of the temperature due to Galactic absorption. The Galactic contribution was estimated from a cylindrically symmetric parameterization of the Galactic emissivity with a frequency dependence. A model for the 3D Galactic electron density was used to estimate the magnitude of the absorption effect.

Different treatments for the frequency dependence of the Galactic emission are included in ULSA, namely a spectral scaling using a single spectral index, a frequency-dependent spectral index, and a direction-dependent spectral index. In this work, we used the latter setting of the model, because we believe this to be more accurate than a single spectral index and we did not observe a significant difference between using a single spectral index or a frequency-dependent index.

The direction-dependent spectral indices were obtained from a fit to the information from nine reference maps between \SI{35}{MHz} and \SI{80}{MHz}, where also a model of the contribution from free-free emission was included. The reference maps at low frequencies are numerous, albeit all except for one map are from the LWA1 survey \citep{Dowell_2017}, which therefore should have a large influence on the model output. A separate spectral index was used for isotropic extragalactic emission. Galactic free-free absorption is treated with special care in ULSA. However, this effect is only relevant at frequencies lower than those that we considered here.

\begin{table*}
	\caption{Overview of the compared radio sky models.}
	\label{tab:sky_models}
	\centering
	\begin{tabular}{c c c c c c c c}
	\hline\hline
	Sky model & Modeling approach & Includes CMB & References \\ \hline
      LFmap  & Spectral scaling & Yes & 1 \\
      GSM (2008)  & PCA & No & 2 \\
      GSM (2016)  & PCA & Yes & 3 \\
      LFSM  & PCA & Yes & 4 \\
      GMOSS  & Physical emission model & No & 5 \\
      SSM  & Spectral scaling + point sources & Yes & 6 \\
      ULSA  & Physical emission model + spectral scaling & Yes & 7 \\
      \hline
	\end{tabular}
        \tablefoot{For the models that do not include the CMB, we added it manually in our comparison.}
        \tablebib{
	(1) \citet{LFmap}; (2) \citet{de_Oliveira_Costa_2008}; (3) \citet{Zheng_2016}; (4) \citet{Dowell_2017}; (5) \citet{GMOSS}; (6) \citet{SSM}; (7) \citet{ULSA}.
	}
\end{table*}

\subsection{Other parametrizations of the Galactic background brightness}
Besides the interpolation models for generating all-sky maps, there are also some parametrizations of the spectrum of the average brightness of the Galactic background. We did not include them in the comparison of the interpolation models but present and show them here for completeness.

One of these parametrizations was introduced by \citet{Cane_1979} and assumes a superposition of Galactic and extragalactic contributions to the brightness including absorption by the Galactic disk and was fitted to a multitude of measurements of the polar regions of the Galaxy. This yields lower brightness temperatures than expected for an average of the whole sky because the bright Galactic center is not considered here. To accommodate for this, a correction factor of ${\sim} 1.3$ to the Galactic brightness contribution was found by \citet{Dulk_2001} and \citet{Duric_2003}. In this work, we show the such-corrected parametrization and refer to it as "Cane."

The parametrization by Cane was revisited again by Tokarsky, Konovalenko, and Yerin \citep{Tokarsky_2017}, where the corrections were summarized and another expression for the average brightness temperature as a function of frequency was given based on results from \citet{Krymkin_1971}. We refer to this parametrization here as "TKY." Both descriptions of the Galactic background brightness are applicable only for the low-frequency radio regime ($<\SI{100}{MHz}$), where the Galaxy is the dominant contribution to the background.

\begin{table*}
	\caption{Summary of all presented reference maps with their sky coverage and quoted uncertainties.}
	\label{tab:reference_maps}
	\centering
	\begin{tabular}{c c c c c c c c}
	\hline\hline
	Map no. & Frequency $\nu$ & Covered declination & $\sigma_k$ & $\sigma_{T_0}$ & $\sigma_{T_0}$ (normalized) & Used in & References \\
	 & (MHz) & & (\%) & (K) & (\%) & & \\ \hline
      1  & 10 & $\SI{-6}{\degr} < \delta < \SI{74}{\degr}$ & 9* & $2\cdot10^4$ & 6.8 & 1,2,3,6 & 1             \\
      2  & 22 & $\SI{-28}{\degr} < \delta < \SI{80}{\degr}$ & 16 & $5\cdot10^3$ & 11.0 & 1,2,3,4,6 & 2               \\
      3  & 35 & $\SI{-40}{\degr} < \delta < \SI{90}{\degr}$ & 20 & 10 & 0.1 & 7 & 3 \\
      4  & 38 & $\SI{-40}{\degr} < \delta < \SI{90}{\degr}$ & 20 & 10 & 0.1 & 7 & 3 \\
      5  & 40 & $\SI{-40}{\degr} < \delta < \SI{90}{\degr}$ & 20 & 10 & 0.1 & 3,7 & 3 \\
      6  & 45 & $\SI{-90}{\degr} < \delta < \SI{65}{\degr}$ & 10/15 & 125$^{\dagger}$ & 1.5 & 1,2,3,(4),6,7 & 4, 5 \\
      7  & 50 & $\SI{-40}{\degr} < \delta < \SI{90}{\degr}$ & 20 & 10 & 0.2 & 3,7 & 3 \\
      8  & 60 & $\SI{-40}{\degr} < \delta < \SI{90}{\degr}$ & 20 & 10 & 0.2 & 3,7 & 3 \\
      9  & 70 & $\SI{-40}{\degr} < \delta < \SI{90}{\degr}$ & 20 & 10 & 0.4 & 3,7 & 3 \\
      10  & 74 & $\SI{-40}{\degr} < \delta < \SI{90}{\degr}$ & 20 & 10 & 0.4 & 7 & 3 \\
      11  & 80 & $\SI{-40}{\degr} < \delta < \SI{90}{\degr}$ & 20 & 10 & 0.5 & 3,7 & 3 \\
      12  & 85 & $\SI{-25}{\degr} < \delta < \SI{25}{\degr}$ & 7 & 120 & 6.8 & 2,6 & 6 \\
      13  & 150 & $\SI{-25}{\degr} < \delta < \SI{25}{\degr}$ & 5 & 40 & 9.4 & 2,5,6 & 6 \\
      14  & 178 & $\SI{-5}{\degr} < \delta < \SI{88}{\degr}$ & 10 & 15 & 5.4 & (1,2,3,6) & 7 \\
      15.a  & 408 & $\SI{-90}{\degr} < \delta < \SI{90}{\degr}$ & 10/5 & 3 & 9.0 & 1,5 & 8 \\
      15.b  & 408 & $\SI{-90}{\degr} < \delta < \SI{90}{\degr}$ & 10/5 & 3 & 9.0 & 4 & 9 \\
      15.c  & 408 & $\SI{-90}{\degr} < \delta < \SI{90}{\degr}$ & 10/5 & 3 & 9.0 & 2,3,6,7 & 10 \\
      \hline
	\end{tabular}
	\tablefoot{The two values for the relative scale uncertainty of map No.\ 4 refer to the two different estimations for this in the publications of the northern and southern part of the survey, respectively. For scale uncertainties with a (*) no explicit value was given by the authors. Instead, we estimated it by taking half of the smallest contour interval of that map and dividing it by its minimum brightness temperature. In the same way, zero-level errors with a $(^{\dagger})$ were estimated by taking half of the smallest contour interval. To give the zero-level errors as a fraction of the average sky temperature, we calculated the latter from Eq.\ \ref{eq:Tsky_average}, where we took the average from using all seven considered sky models. For indicating in which models the maps are used, we use the following notation: 1 = GSM, 2 = GSM16, 3 = LFSM, 4 = LFmap, 5 = GMOSS, 6 = SSM, 7 = ULSA. Numbers in parentheses mark that the map is only used indirectly in the respective model.}
	\tablebib{
	(1) \citet{Caswell_1976}; (2) \citet{Roger_1999}; (3) \citet{Dowell_2017}; (4) \citet{Alvarez_1997}; (5) \citet{Maeda_1999}; (6) \citet{Landecker_1970}; (7) \citet{Turtle_1962}; (8) \citet{Haslam_1982}; (9) \citet{Platania_2003}; (10) \citet{Remazeilles_2015}.
	}
\end{table*}

Furthermore for comparison purposes, we show the simple approach of scaling a full-sky map with just one single spectral index. As the reference map, we used the \SI{408}{MHz} map by \citet{Haslam_1982} as improved by \citet{Remazeilles_2015}. Here, we refer to this description as "Haslam," but also did not include it in the comparison of the interpolation models because of its drawbacks. Using a single spectral index does not represent reality, as the spectral index varies for different regions of the sky and different frequency regimes \citep{Mozdzen_2019, Dickinson_2019}. Although often used in studies of fast radio bursts, the Haslam description should not be seen as equivalent to the more sophisticated interpolation models \citep{Price_2021}. We show the Haslam results for a spectral index between $-2.62$ and $-2.46$.

\section{Reference maps}\label{sec:ReferenceMaps} 
\label{sec:3_Maps}
The presented radio sky interpolation models rely on a number of reference maps. Consequently, the accuracy of the models is dependent on the accuracy of these maps and their inherited uncertainties. Therefore, in the following, we give an overview of the reference maps used in the considered interpolation models.

\begin{figure*}
  \begin{subfigure}[c]{0.33\hsize}
    \center
    \includegraphics[width=1.\hsize]{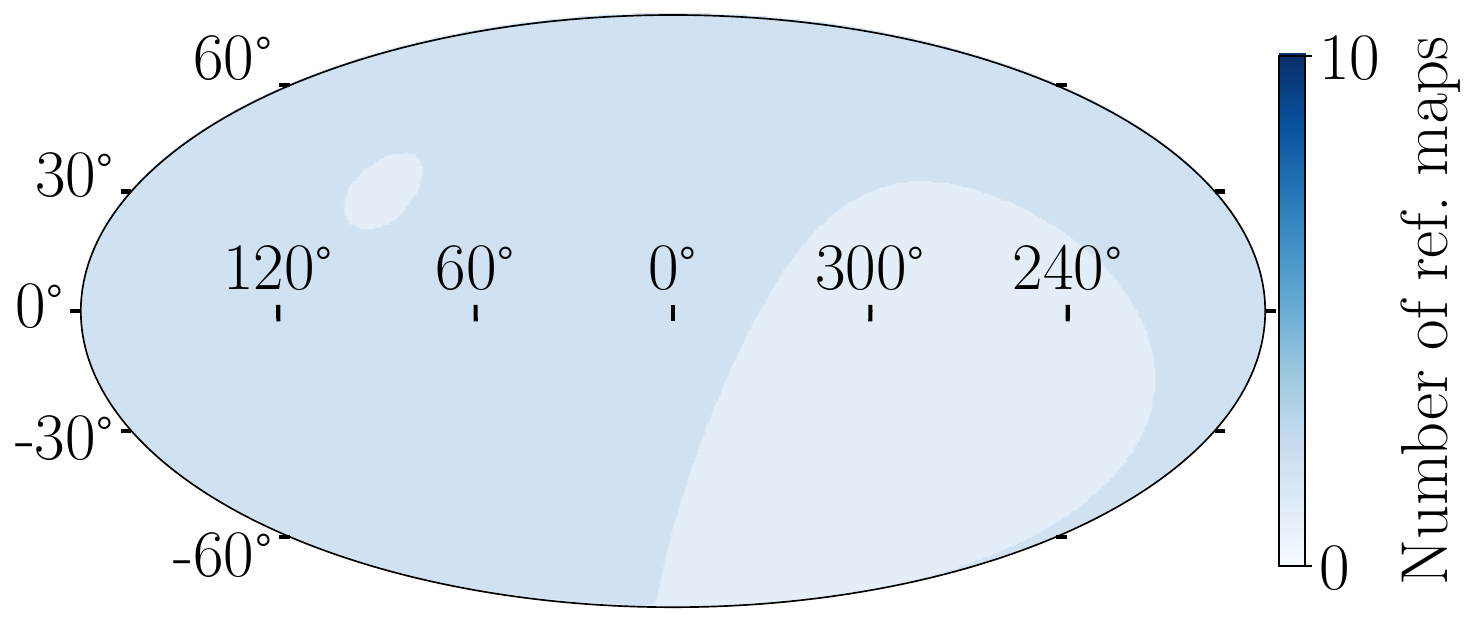}
    \subcaption{\;\;\;\;\;\;}
  \end{subfigure}
  \begin{subfigure}[c]{0.33\hsize}
    \center
    \includegraphics[width=1.\hsize]{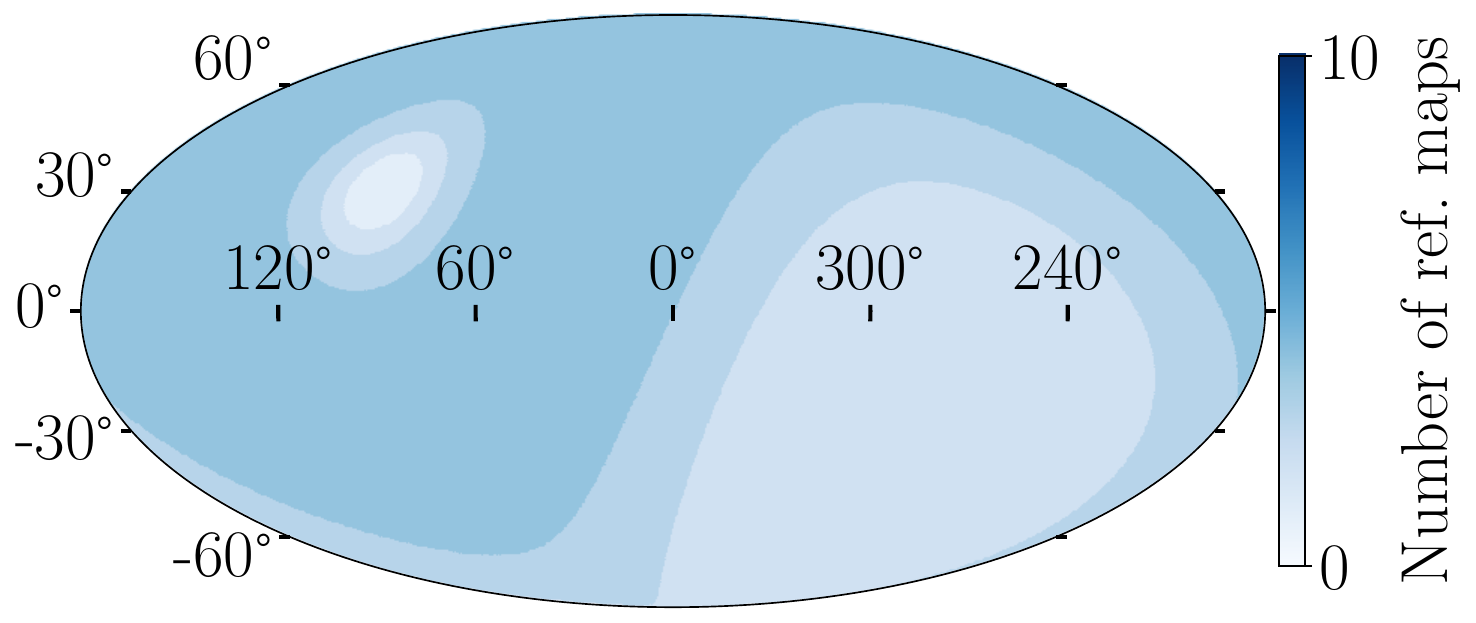}
    \subcaption{\;\;\;\;\;\;}
  \end{subfigure}
  \begin{subfigure}[c]{0.33\hsize}
    \center
    \includegraphics[width=1.\hsize]{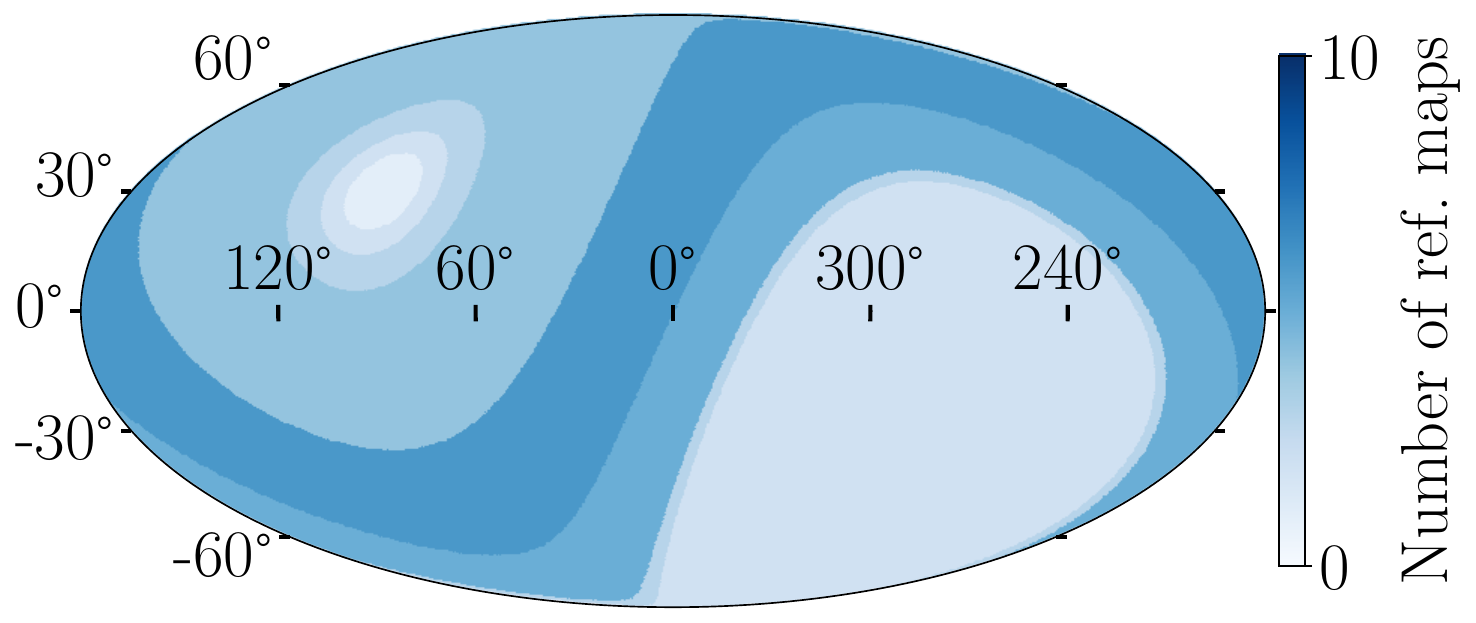}
    \subcaption{\;\;\;\;\;\;}
  \end{subfigure}
  \\
  \hspace*{\fill}
  \begin{subfigure}[c]{0.33\hsize}
    \center
    \includegraphics[width=1.\hsize]{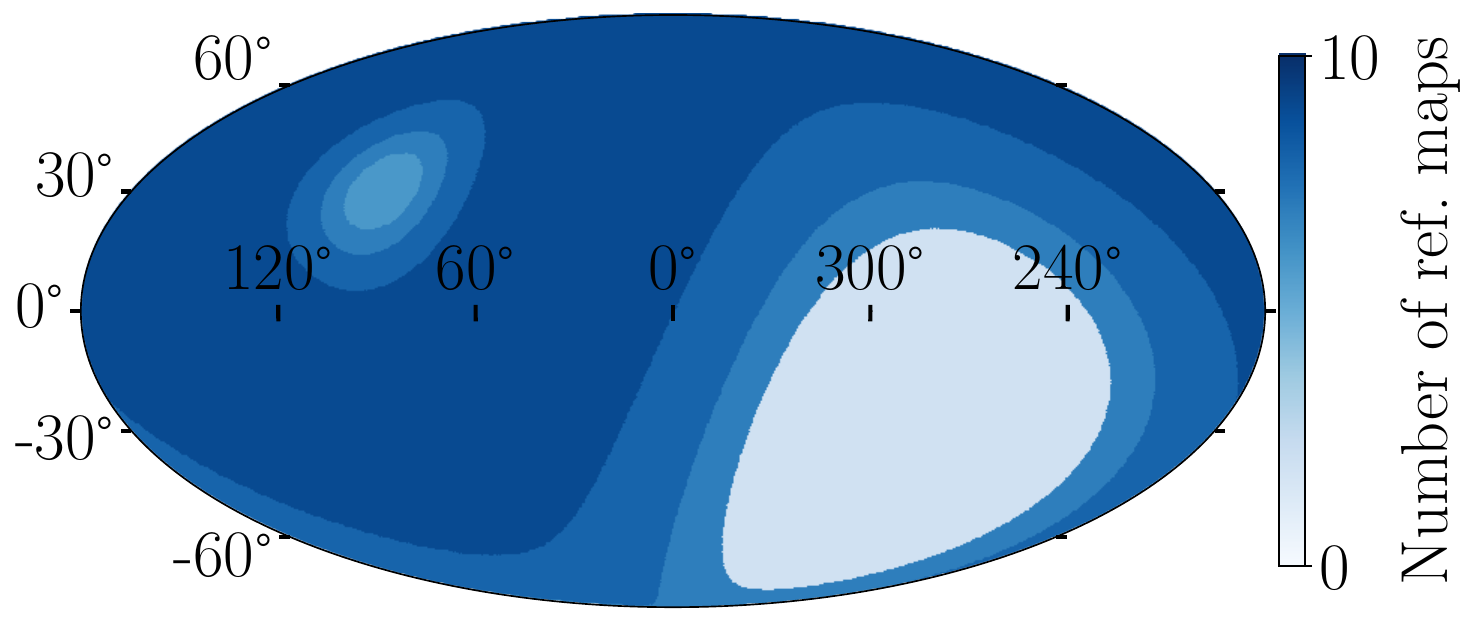}
    \subcaption{\;\;\;\;\;\;}
  \end{subfigure}
  \hspace*{\fill}
  \begin{subfigure}[c]{0.33\hsize}
    \center
    \includegraphics[width=1.\hsize]{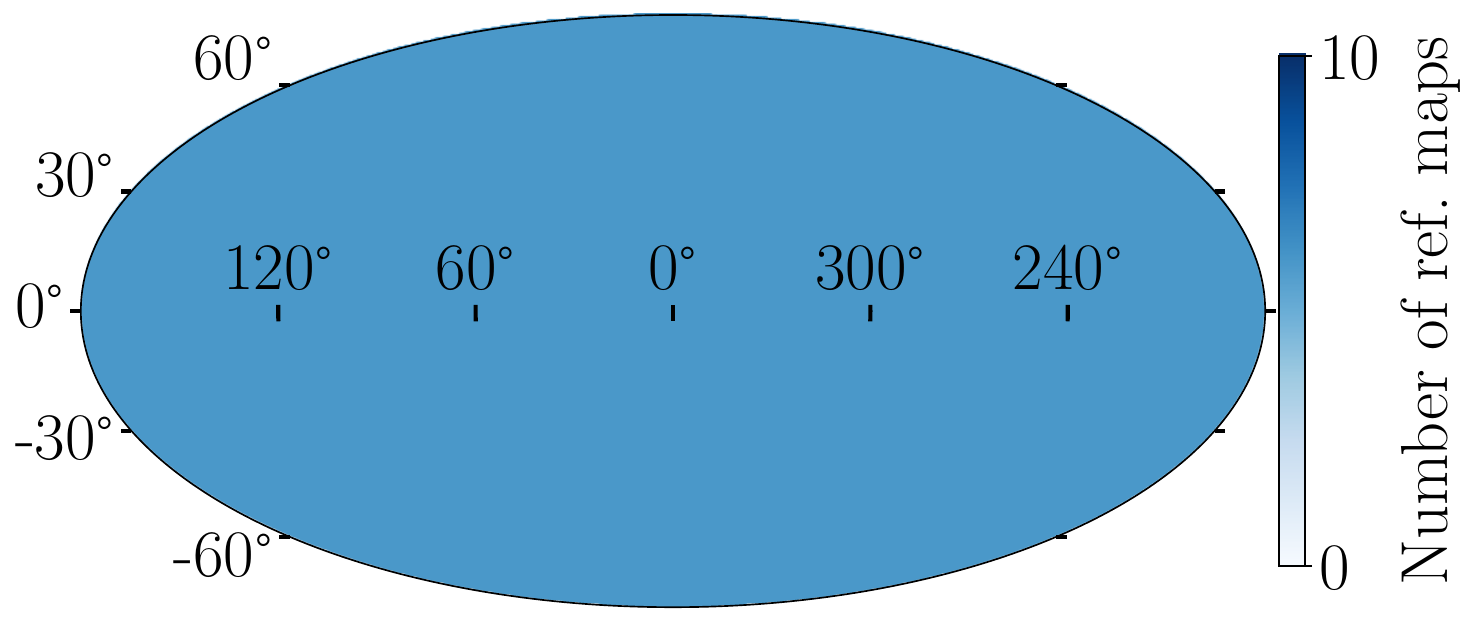}
    \subcaption{\;\;\;\;\;\;}
  \end{subfigure}
  \hspace*{\fill}
  \\
  \hspace*{\fill}
  \begin{subfigure}[c]{0.33\hsize}
    \center
    \includegraphics[width=1.\hsize]{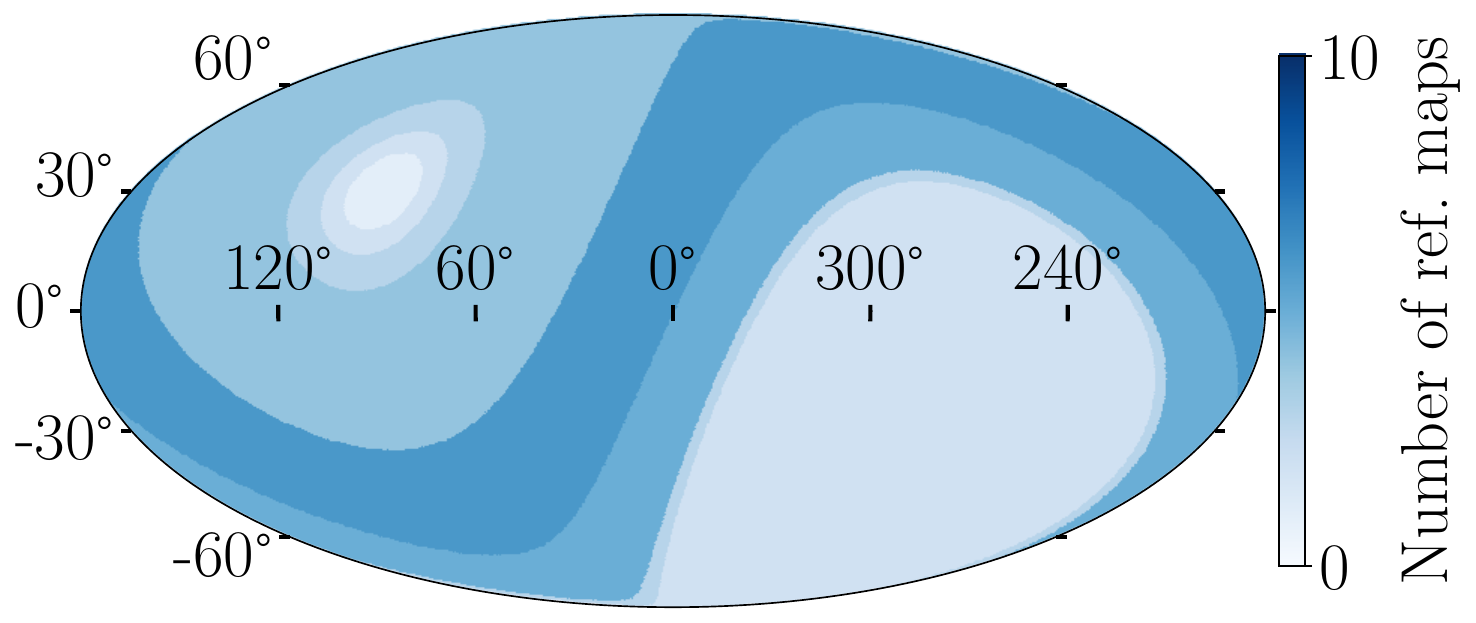}
    \subcaption{\;\;\;\;\;\;}
  \end{subfigure}
  \hspace*{\fill}
  \begin{subfigure}[c]{0.33\hsize}
    \center
    \includegraphics[width=1.\hsize]{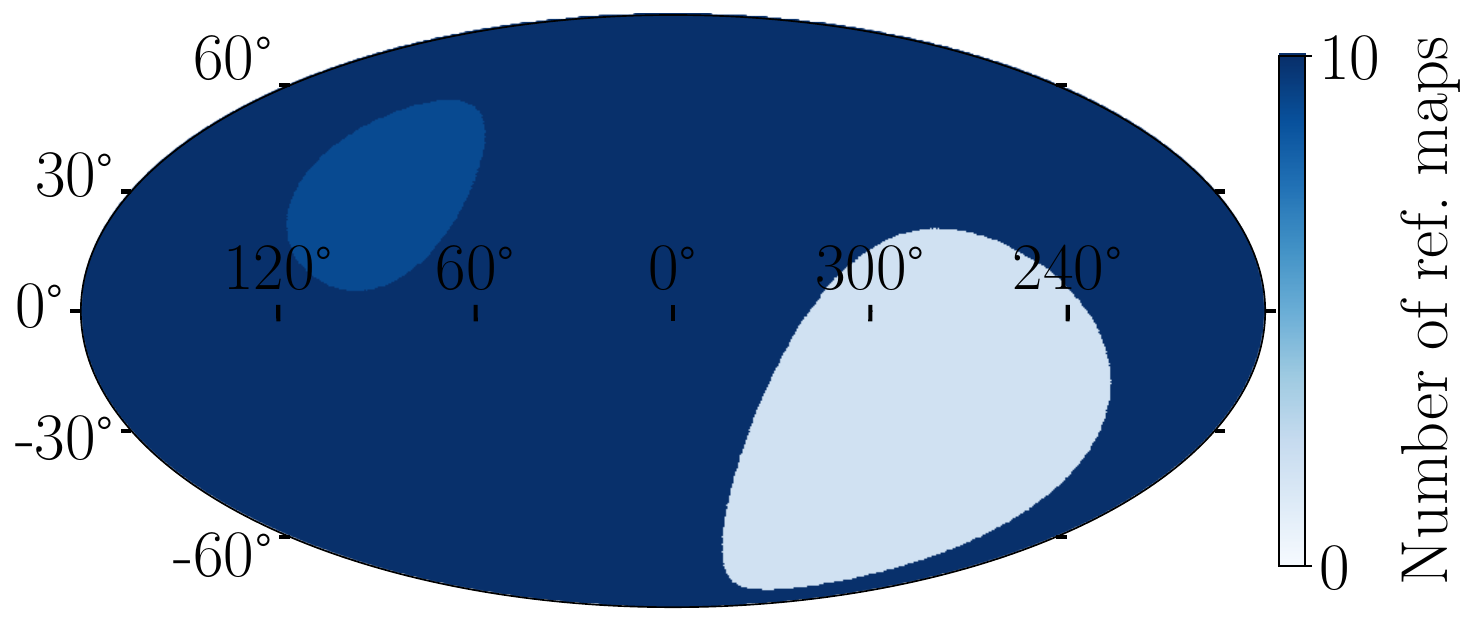}
    \subcaption{\;\;\;\;\;\;}
  \end{subfigure}
  \hspace*{\fill}
  \caption{Sky coverages of the used reference maps for each sky model laid on top of each other and plotted in Galactic coordinates. The models are numbered (a) LFmap, (b) GSM, (c) GSM16, (d) LFSM, (e) GMOSS, (f) SSM, (g) ULSA. For regions of lighter color fewer measurements from reference maps are available than for regions of darker color.}
  \label{fig:Coverage_maps}
\end{figure*}

\paragraph{\SI{10}{MHz} \normalfont{\citep{Caswell_1976}:}}
This map was obtained using the 10-MHz array at the Dominion Radio Astrophysical Radio Observatory (DRAO) in Canada. The observations were done during winter nights to minimize ionospheric influence. Thus, the sky coverage is limited in right ascension from $0^\mathrm{h}$ to $16^\mathrm{h}$. Relative calibration of the received signal strength to sky brightness temperature was performed by comparing the observations to a published \SI{178}{MHz} map from \citet{Turtle_1962}, that was scaled using a power-law relation as in Eq.\ \ref{eq:power_law} and spectral indices of \num{-2.40} and \num{-2.75} for the Galactic and extragalactic isotropic emission, respectively. The sky brightness is mapped in contours indicating specific temperature levels. The contour interval between these levels is \SI{2e4}{K}.

\paragraph{\SI{22}{MHz} \normalfont{\citep{Roger_1999}:}}
The sky map at \SI{22}{MHz} was also produced from measurements done at the DRAO. The applied antenna gain is based on an assumed value for the flux density of Cygnus A \citep{Roger_1969}. A cross-check of the map with the \SI{408}{MHz} map by \citet{Haslam_1982} was performed to compare the temperature scales by producing scatter plots of the brightness temperatures ($T$-$T$ plots) from both maps for different declinations from the zenith. Then, the temperature scale of the \SI{22}{MHz} map was tuned to overall match the ratio at the zenith, for which the authors believed their instrument to be understood better. The $T$-$T$ plot for the zenith direction showed no discrepancy with the zero-level of the temperature scale. Also, the differential spectral index derived from the $T$-$T$ plot between the \SI{22}{MHz} map and the \SI{408}{MHz} map was in agreement with other measurements.

\paragraph{\normalfont{LFSS} \SI{35}{MHz}, \SI{38}{MHz}, \SI{40}{MHz}, \SI{50}{MHz}, \SI{60}{MHz}, \SI{70}{MHz}, \SI{74}{MHz}, and \SI{80}{MHz} \normalfont{\citep{Dowell_2017}:}} Within the LFSS at the LWA1, sky maps were produced at a variety of frequencies. The gain of the LWA1 antennas was derived from an electromagnetic antenna model in combination with a correction using bright pulsars at different elevations as reference sources. Previous observations of the flux density of Cygnus A by \citet{Baars_1977} were used to convert the measurements into sky temperature. A temperature calibration system was used to achieve absolute accuracy of \SI{10}{K} or better. 

\paragraph{\SI{45}{MHz} \normalfont{\citep{Alvarez_1997, Maeda_1999}:}}
At \SI{45}{MHz}, there are two separate sky surveys available, one covering the northern hemisphere and one covering the southern hemisphere, conducted at the Middle and Upper Atmosphere Radar in Japan and the Maipu Radio Astronomy Observatory in Chile, respectively. Calibration of the temperature scale of the latter survey was performed by using a spectral interpolation from data in a well-observed reference region and comparison with the measurements. The map was later used to calibrate the map of the northern hemisphere as both cover the region from \SI{5}{\degr} to \SI{19}{\degr} in declination.

\paragraph{\SI{85}{MHz}, \SI{150}{MHz} \normalfont{\citep{Landecker_1970}:}}
Surveys at \SI{85}{MHz} and \SI{150}{MHz} were made using the Parkes \SI{64}{m} telescope in Australia. Calibration of the temperature scale was done after every scan with well-matched noise diodes that could be connected to the receivers of the telescope.

\paragraph{\SI{408}{MHz} \normalfont{\citep{Haslam_1982}:}}
Multiple sites (Jodrell Bank MkI(A) telescope in England, Effelsberg \SI{100}{m} telescope in Germany, and Parkes \SI{64}{m} telescope in Australia) were used to observe the sky at \SI{408}{MHz}. The originally assembled map by Haslam et al.\ was calibrated with reference to a sky survey at \SI{404}{MHz} from \citet{Pauliny_1962}. \citet{Remazeilles_2015} believe that the brightness temperature scale is more accurate (${\sim}\SI{5}{\%}$) than originally quoted (${\sim}\SI{10}{\%}$). Because of its importance in the field of radio astronomy, the map and its raw data were restudied multiple times, including destriping it and removing point-like sources \citep{Platania_2003, Remazeilles_2015}.

\paragraph{\normalfont{Summary:}}
The reference maps are listed in Table \ref{tab:reference_maps} together with their quoted uncertainties on the temperature scale, which are typically described by a linear relation. The true temperature $T_\text{true}$ is assumed to relate to the observed temperature $T_\text{obs}$ as

\begin{equation}
    T_\text{true} = k T_\text{obs} + T_0.
\end{equation}

\begin{figure*}
  \begin{subfigure}[c]{0.33\hsize}
    \center
    \includegraphics[width=1.\hsize]{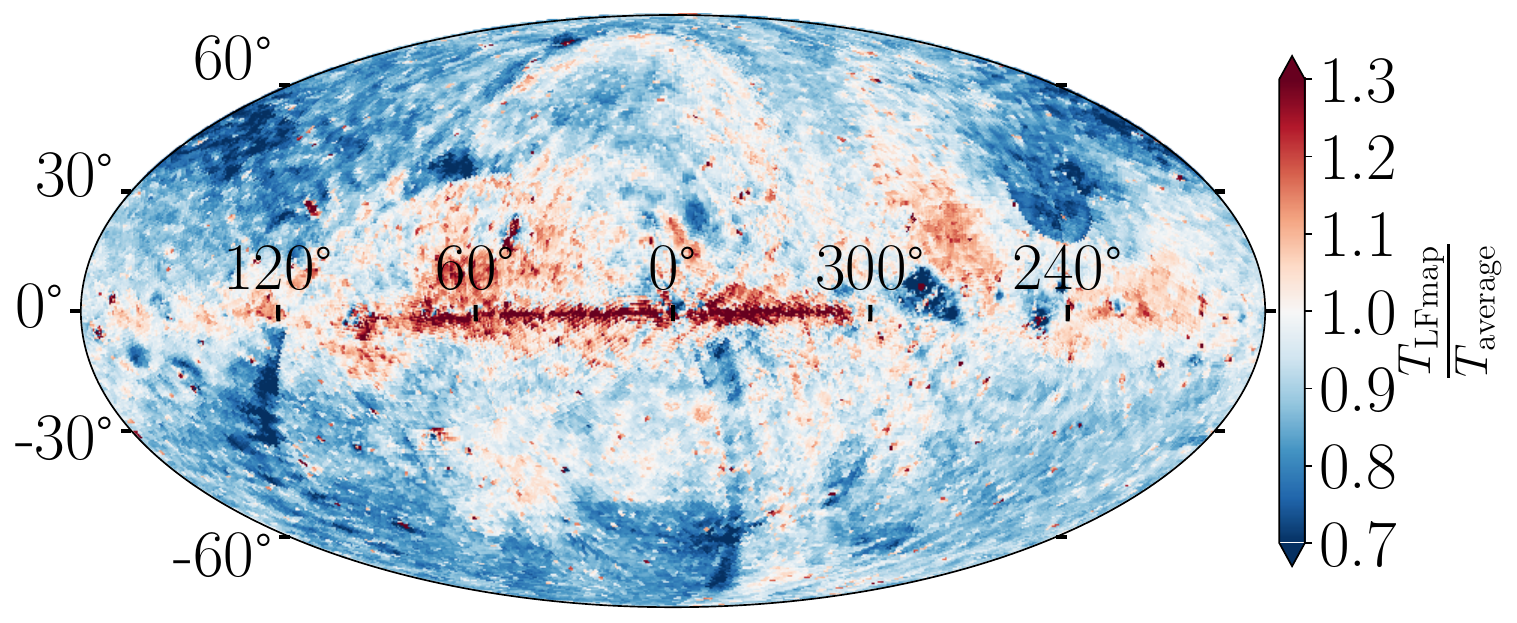}
    \subcaption{\;\;\;\;\;\;}
  \end{subfigure}
  \begin{subfigure}[c]{0.33\hsize}
    \center
    \includegraphics[width=1.\hsize]{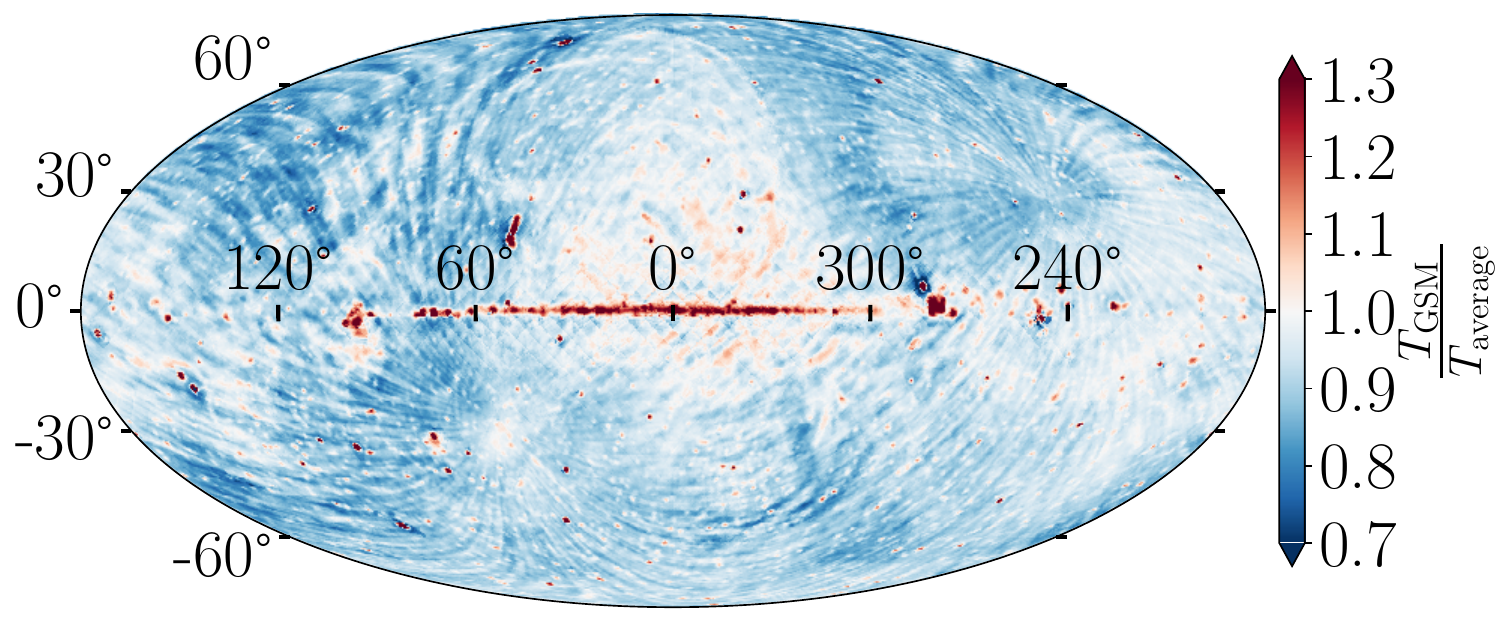}
    \subcaption{\;\;\;\;\;\;}
  \end{subfigure}
  \begin{subfigure}[c]{0.33\hsize}
    \center
    \includegraphics[width=1.\hsize]{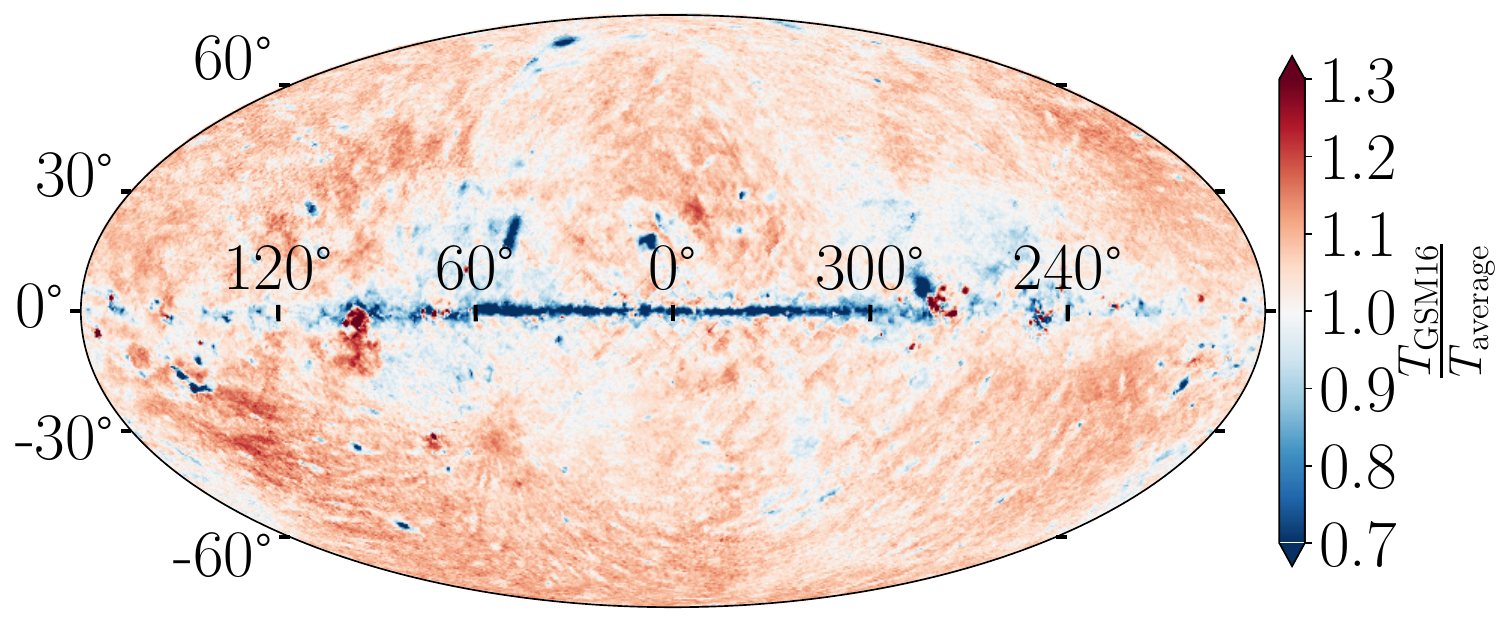}
    \subcaption{\;\;\;\;\;\;}
  \end{subfigure}
  \\
  \hspace*{\fill}
  \begin{subfigure}[c]{0.33\hsize}
    \center
    \includegraphics[width=1.\hsize]{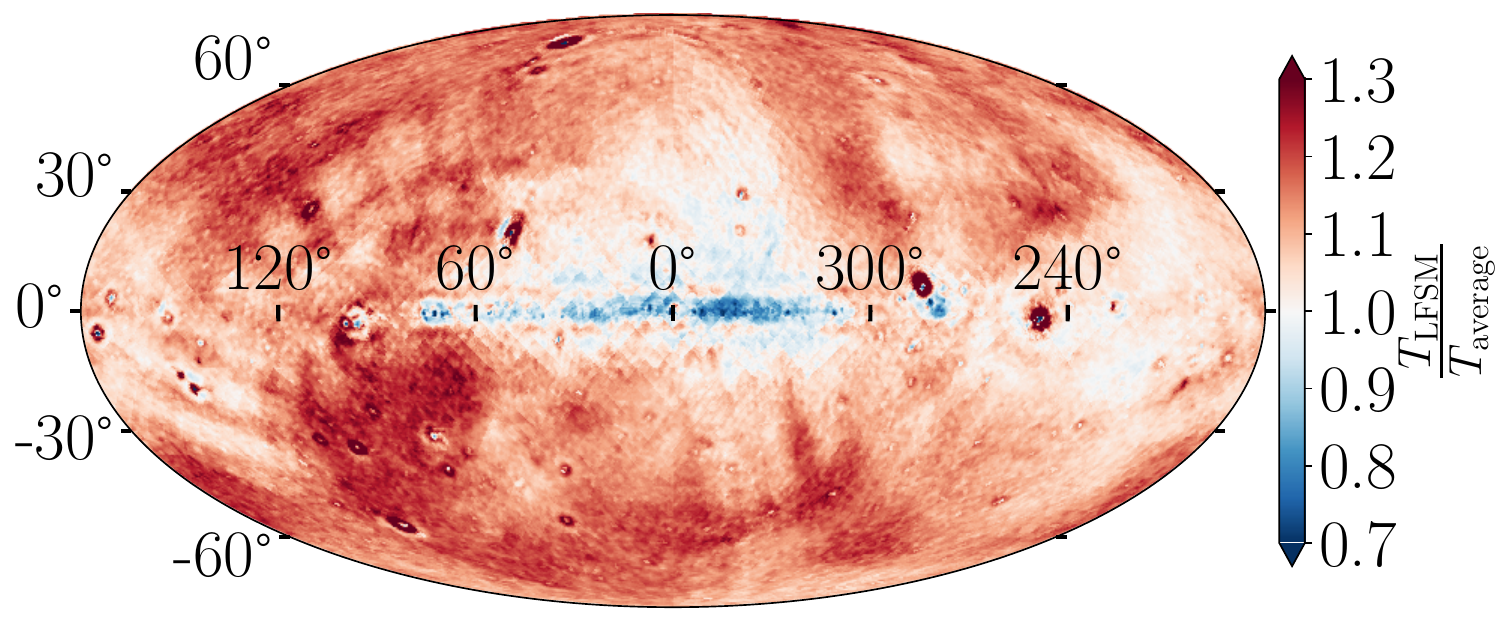}
    \subcaption{\;\;\;\;\;\;}
  \end{subfigure}
  \hspace*{\fill}
  \begin{subfigure}[c]{0.33\hsize}
    \center
    \includegraphics[width=1.\hsize]{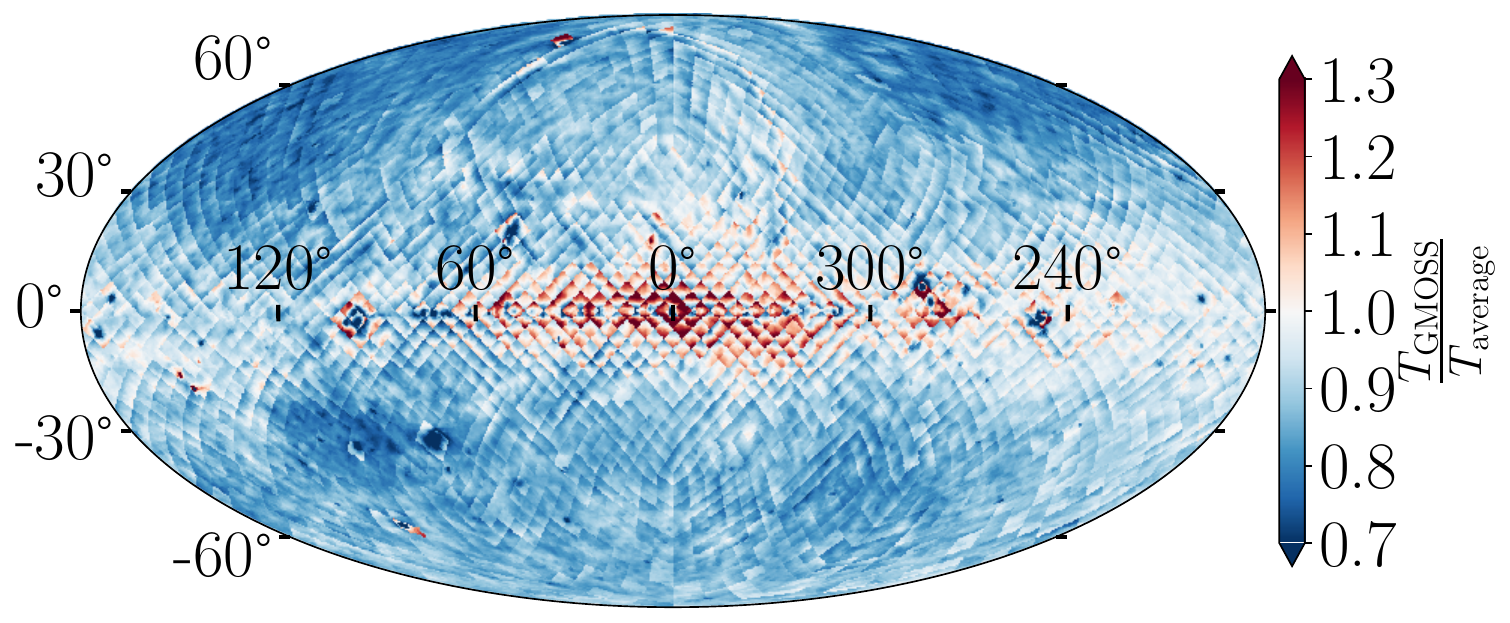}
    \subcaption{\;\;\;\;\;\;}
  \end{subfigure}
  \hspace*{\fill}
  \\
  \hspace*{\fill}
  \begin{subfigure}[c]{0.33\hsize}
    \center
    \includegraphics[width=1.\hsize]{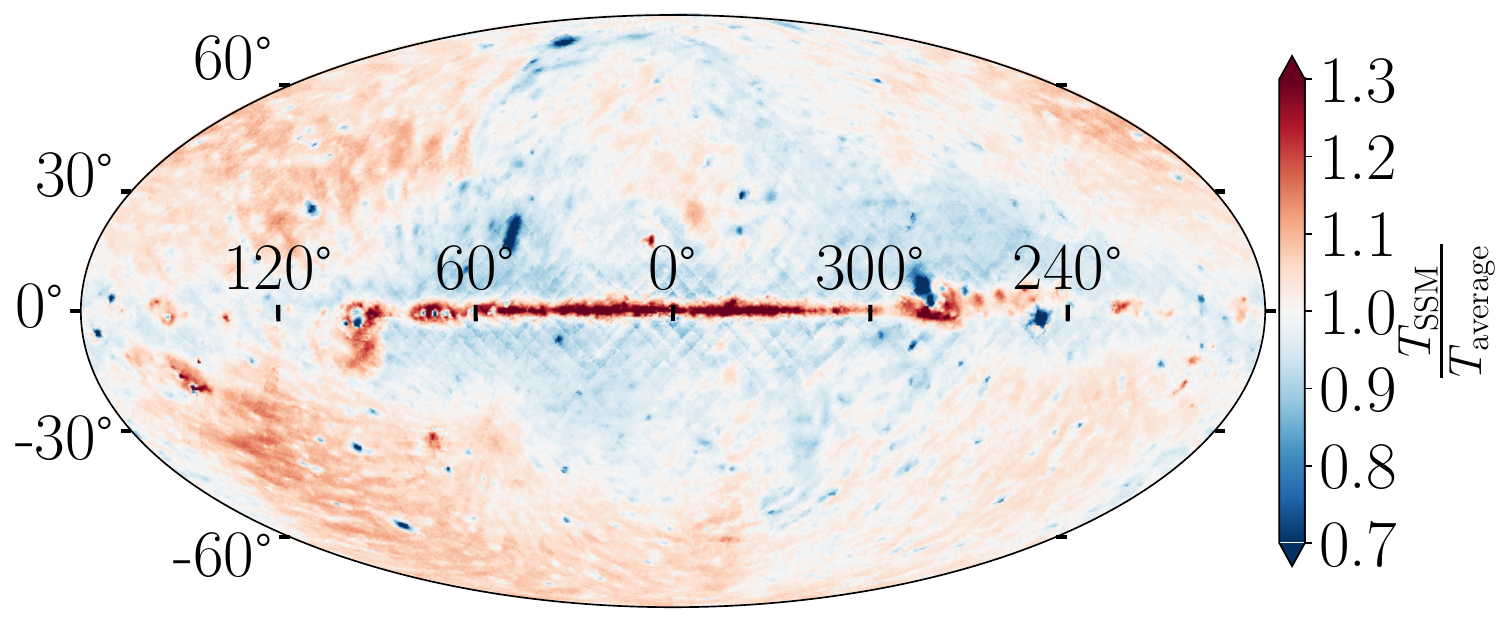}
    \subcaption{\;\;\;\;\;\;}
  \end{subfigure}
  \hspace*{\fill}
  \begin{subfigure}[c]{0.33\hsize}
    \center
    \includegraphics[width=1.\hsize]{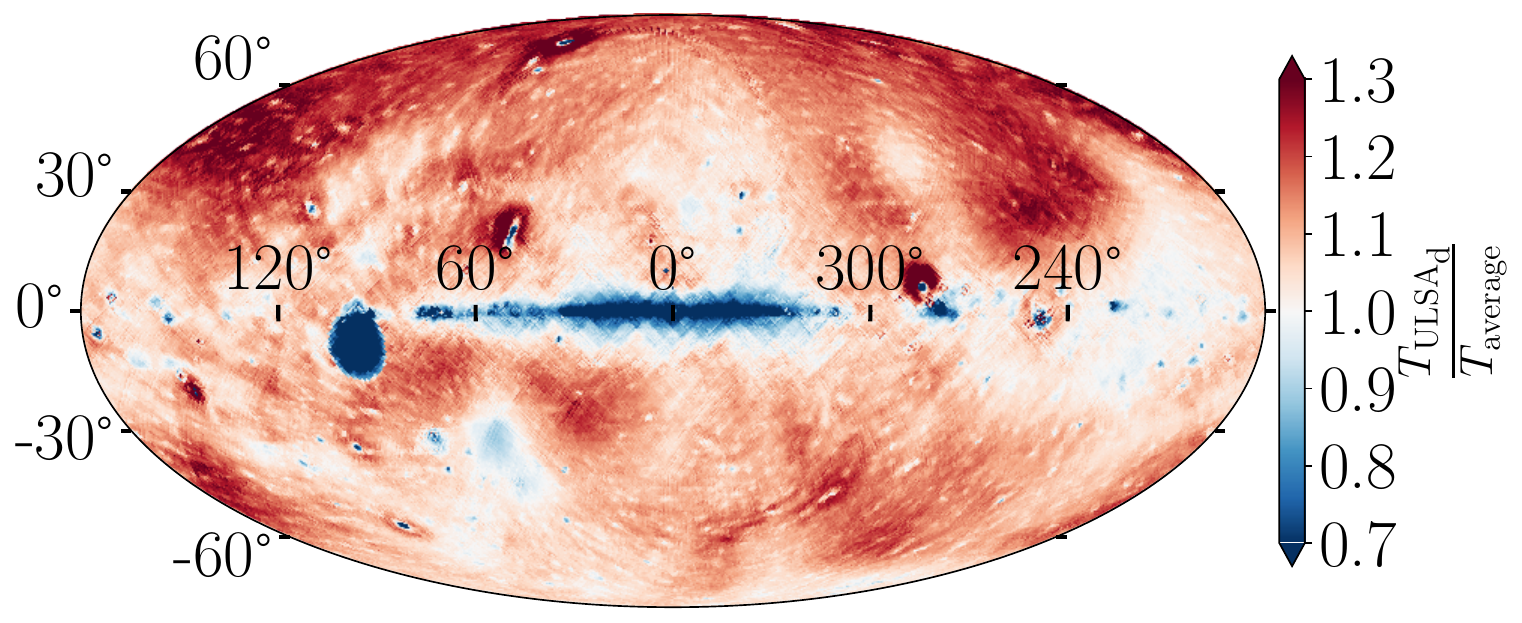}
    \subcaption{\;\;\;\;\;\;}
  \end{subfigure}
  \hspace*{\fill}
  \caption{Sky maps showing the temperature ratio of each model to the average from all seven models at \SI{50}{MHz} in Galactic coordinates. The models are denoted as (a) LFmap, (b) GSM, (c) GSM16, (d) LFSM, (e) GMOSS, (f) SSM, (g) ULSA.}
  \label{fig:Ratio maps}
\end{figure*}

The scaling factor $k$ ideally has a value of 1, but it inherits the relative scale uncertainty $\sigma_k$, which goes up to 20\% for the reference maps presented here, while the largest uncertainties are quoted for the results of the Low Frequency Sky Survey (LFSS) (map No.\ \numrange{3}{5}, \numrange{7}{11}). The zero-level $T_0$ represents an absolute offset of the temperature scale and deviations from the ideal case ($T_0=\SI{0}{K}$) are reflected in the zero-level error $\sigma_{T_0}$. The zero-level errors of the reference maps are given in Table \ref{tab:reference_maps} as absolute values and also relative to the average brightness temperature of the whole sky at the respective frequency, where the average sky brightness temperature is calculated using Eq.\ \ref{eq:Tsky_average}. The relative zero-level errors amount up to $\sim 11\%$, while they are negligible for the LFSS results.

Furthermore, Table \ref{tab:reference_maps} gives the information, in which models each reference map is used. This information is also visualized in Fig.\ \ref{fig:Coverage_maps}, where for each model the covered sky regions of all used reference maps are superimposed. LFmap relies on only two reference maps with the approach of spectral scaling. ULSA uses the most reference maps with a maximum of ten maps covering the region between \SI{-40}{\degr} and \SI{65}{\degr} of declination. However, there is a lack of available reference maps for all models around the south celestial pole. 

\paragraph{\normalfont{Corrections to the maps at} \SI{45}{MHz}, \SI{150}{MHz}, and \SI{408}{MHz}:}
The sky surveys at \SI{45}{MHz} that are used in the interpolation models considered here were done separately for the southern \citep{Alvarez_1997} and northern hemisphere \citep{Maeda_1999}. In a later study, the maps were combined into an all-sky survey with a correction of the zero-level of \SI{-544}{K} \citep{Guzman_2011} to the combined map. This corresponds to ${\sim}\SI{6.5}{\%}$ of the average sky temperature at that frequency and is more than three times larger than the originally quoted error on the zero-level. In the same study, also a zero-level correction for the original \SI{408}{MHz} map of \SI{-3.46}{K} was determined, which corresponds to ${\sim}\SI{10.4}{\%}$ of the average sky temperature.

Another recalibration was performed for the combined \SI{45}{MHz} map (although without the mentioned zero-level correction) and for the \SI{150}{MHz} map \citep{Monsalve_2021}. There, the maps were corrected for temperature scale and zero-level to best match data taken with the Experiment to Detect the Global EoR (epoch of reionization) Signature (EDGES) \citep{Bowman_2018}. Scale correction factors are \SI{1.076\pm 0.017}{} and \SI{1.112\pm 0.012}{} for the sky temperature of the \SI{45}{MHz} and \SI{150}{MHz} maps, respectively. Zero-level corrections for the surveys are \SI{-160\pm 78}{K} and \SI{0.7\pm 6.0}{K}, which correspond to $\SI{1.9\pm 0.9}{\%}$ and $\SI{0.2\pm 1.4}{\%}$ of the average sky temperature at the respective frequencies.

Corrections to the temperature scale and zero-level by matching to other, sometimes newer data can be bigger than originally quoted uncertainties. This hints towards an underestimation of the latter and poses a challenge when trying to place trust on individual surveys.
\\

Further maps at higher frequencies are used in the interpolation models as well. Many of them were generated from data taken with the space-based instruments Wilkinson Microwave Anisotropy Probe (WMAP) and Planck \citep{Hinshaw_2009,Planck_2015}. These maps are important for modeling sky regions where fewer low-frequency surveys were performed, but they are not discussed in more detail here.

\section{Comparison of the sky model predictions} 

\label{sec:4_Comparison}
Besides the quoted accuracies of the reference maps, which propagate into the interpolation models, we compared the output of the models directly. We studied the deviations between them and determined their level of agreement, which we use as an estimator for the systematic uncertainty in predicting the diffuse Galactic radio emission on an absolute scale.

The \texttt{PyGDSM} \citep{PyGDSM} package was used as an interface to GSM, GSM16, and LFSM. It employs the \texttt{healpy} \citep{healpy} package to provide the temperature map output of the models in the \texttt{HEALPix} \citep{healpix} format. The source codes of LFmap, GMOSS, SSM, and ULSA are all separately available at their authors' respective websites\footnote{LFmap: \url{https://www.astro.umd.edu/~emilp/LFmap}\\ GMOSS: \url{https://github.com/mayurisrao/GMOSS}\\ SSM: \url{http://tianlai.bao.ac.cn/~huangqizhi}\\ ULSA: \url{https://github.com/Yanping-Cong/ULSA}}. After generating the output maps using these four models, we converted them as well into the \texttt{HEALPix} format. A software toolkit for the Galactic calibration, which also includes a unified, simpler access to all considered models, is under development \citep{Fodran_2023}.

Exemplarily, sky maps are shown in Fig.\ \ref{fig:Ratio maps} which display the temperature ratio between the output of each model and the average at \SI{50}{MHz}. The average is calculated pixelwise from all seven models at the same frequency. At \SI{50}{MHz} LFmap, GSM, GMOSS, and SSM predict a hotter Galactic center than the other three models, while away from the center their predictions are colder, or close to average in the case of SSM. By this depiction, also spatial structures of the differences between the models are visible, for example, some stripe-like features for GSM. In this work, we considered an application of these models for radio antennas with a rather broad beam width, so that spatial structures are averaged out. Therefore, we did not investigate these structures more deeply.

In this study, the comparison was conducted for the frequency range from \num{30} to \SI{408}{MHz}. Typically, radio arrays for the detection of cosmic particles are not operated below \SI{30}{MHz} because of the presence of strong atmospheric noise as we discuss in Sec.\ \ref{sec:5_Ionosphere}. The upper bound arises from limitations within the LFmap and LFSM models.

\subsection{Comparison of the total sky}

\begin{figure}
	\resizebox{\hsize}{!}{\includegraphics{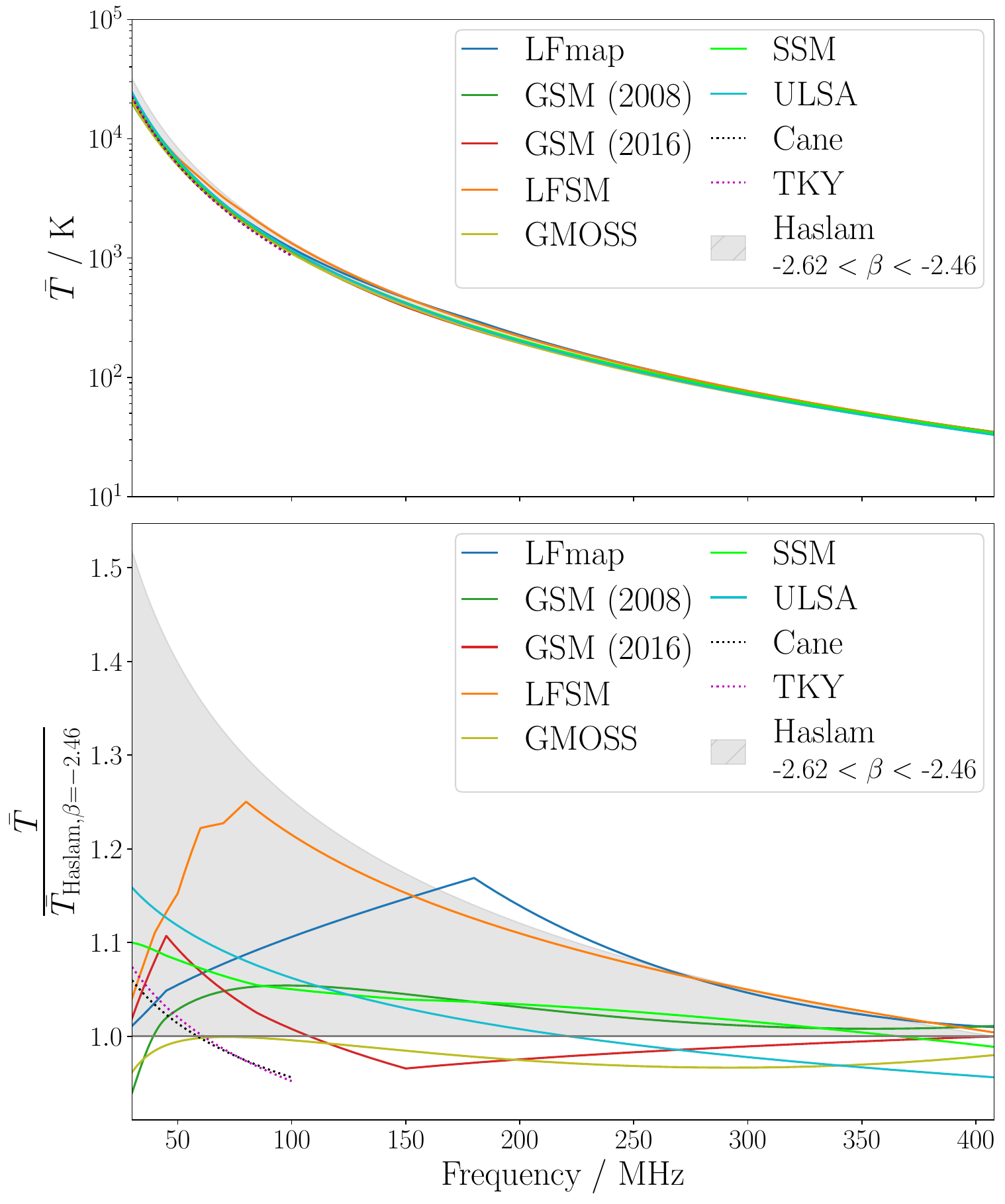}}
	\caption{Average sky temperature as a function of the frequency is plotted for the interpolation models as solid lines and for the parameterizations Cane and TKY as dashed lines. The gray band shows the results for the \SI{408}{MHz} Haslam map when scaled down with a spectral index $\beta$ between $-2.62$ and $-2.46$. The upper plot shows the absolute values, while in the lower plot the results are normalized to those for the scaled Haslam map and a spectral index $\beta=-2.46$.}
	\label{fig:All_models_combined}
\end{figure}

The average temperature of the sky at a given frequency $\nu$ is calculated as

\begin{equation}
\label{eq:Tsky_average}
\bar{T}(\nu) = \frac{1}{4\pi} \int_{-\pi}^{\pi} \mathrm{d} \ell \int_{\frac{-\pi}{2}}^{\frac{\pi}{2}} \mathrm{d} b \cos{(b)}\; T(\nu; \ell, b),
\end{equation}

where $\ell$ is the Galactic longitude and $b$ is the Galactic latitude. The brightness temperature at a specific location in the sky $T(\nu; \ell, b)$ is taken from maps in Galactic coordinates produced with the sky models. The observable $\bar{T}(\nu)$ is used for the first part of the comparison. It gives compressed information of a sky map while ignoring spatial structures. For the Galactic calibration of an antenna without a narrow beam, taking into account fine spatial structures is not as relevant, because the measured quantity of received power is a folding of the whole visible sky through the antenna pattern.

The upper plot in Fig.\ \ref{fig:All_models_combined} shows the average sky temperature as a function of the frequency for the interpolation models in solid lines from $\SI{30}{MHz}$ to $\SI{408}{MHz}$. The gray band was obtained from the Haslam description of the sky brightness by using a spectral index $\beta$ between \num{-2.62} and \num{-2.46} and is shown for comparison. The range for the spectral index was deduced from recent measurements \citep{Mozdzen_2016, Mozdzen_2019} around the relevant frequencies. Furthermore, the previously introduced parametrizations of the average sky brightness below \SI{100}{MHz}, Cane and TKY, are shown for comparison as black and magenta dotted lines, respectively.

The comparison shows that the interpolation models agree well in their shapes over the whole frequency range, while they are systematically shifted relative to each other. This can be seen more quantitatively in the lower plot of Fig.\ \ref{fig:All_models_combined}, in which the data are normalized to the Haslam results with a spectral index $\beta=-2.46$, to better show spectral behavior of the models.

\begin{figure}
	\centering
	\resizebox{\hsize}{!}{\includegraphics{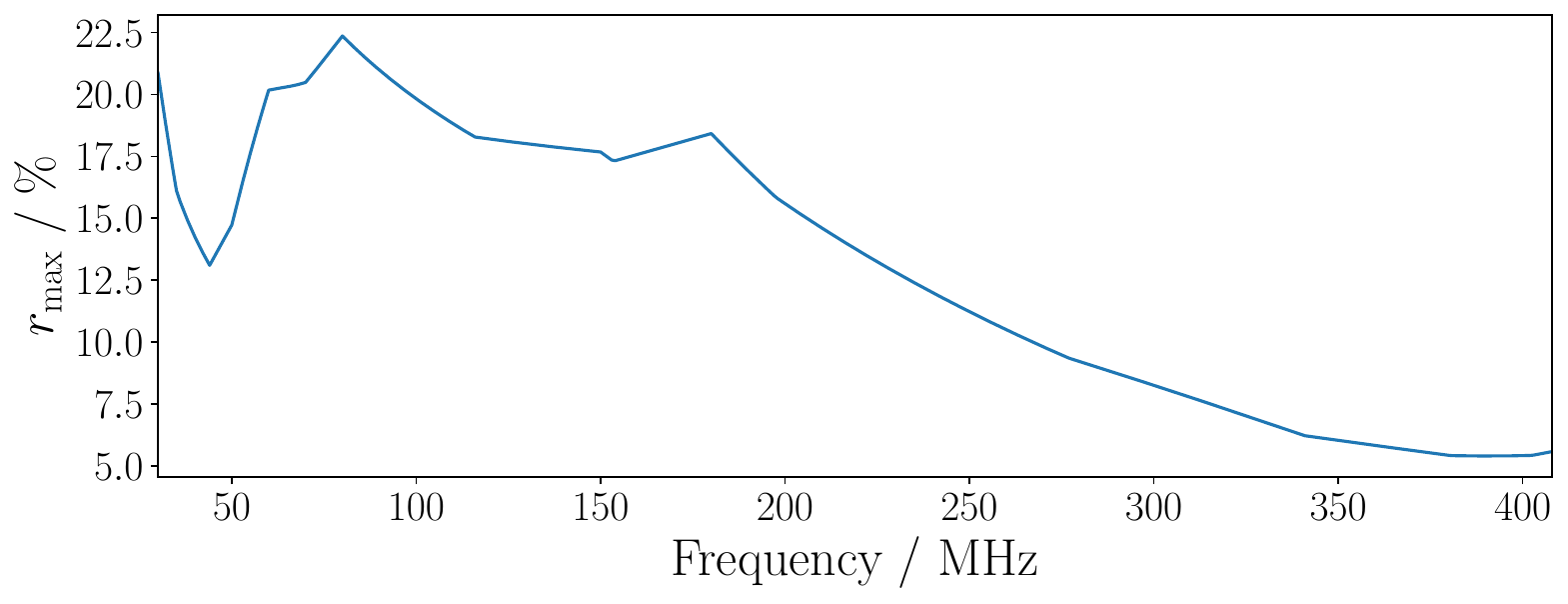}}
	\caption{Maximum relative difference of the average sky temperature between any two of the considered models as a function of frequency. The maximum relative difference is calculated as in Eq.\ \ref{eq:LoA}.}
	\label{fig:All_models_LoA}
\end{figure}

We subsequently calculated the relative difference of the average sky temperatures at a given frequency for any combination of models m\textsubscript{1} and m\textsubscript{2}. From that, we evaluated the maximum relative difference

\begin{equation}
    \label{eq:LoA}
    r_\text{max}(\nu) = \max|2\frac{\bar{T}_{\text{m}_1}(\nu) - \bar{T}_{\text{m}_2}(\nu)}{\bar{T}_{\text{m}_1}(\nu) + \bar{T}_{\text{m}_2}(\nu)}| \text{,}
\end{equation}

which we plot as a function of the frequency, shown in Fig.\ \ref{fig:All_models_LoA}. The maximum relative difference is around 20\% for the lowest frequencies and drops to values around 7\% for frequencies larger than ${\sim}\SI{200}{MHz}$. 

To gauge the level of agreement between the models more concisely, for each of them we integrated the average sky temperature over the frequency range from 30 to 408\,MHz, and calculated a relative difference $r_{\text{m}_1,\text{m}_2}$ pairwise as

\begin{equation}
r_{\text{m}_1,\text{m}_2} = 2 \frac{\int_{30\text{MHz}}^{408\text{MHz}} \bar{T}_{\text{m}_1}(\nu) - \bar{T}_{\text{m}_2}(\nu) \; \text{d} \nu }{\int_{30\text{MHz}}^{408\text{MHz}} \bar{T}_{\text{m}_1}(\nu) + \bar{T}_{\text{m}_2}(\nu) \; \text{d} \nu } \text{,}
\end{equation}

again with any combination of models m\textsubscript{1} and m\textsubscript{2}. The results are listed in Table \ref{tab:model_comparison}. By this comparison, all models agree with each other at a level of 14.3\% or better. We observe a rather homogeneous distribution of the models in terms of the average sky brightness that they predict. There is no aggregation of models in the results of this comparison and, in particular, there is no correlation of the average brightness of the model's prediction with the modeling approach. For example, when looking at the models that use a PCA, LFSM on average predicts the brightest skies while GSM predicts relatively faint skies. The prediction from GSM16 comes out in between. We conclude that the modeling approach is not the dominant reason for differences between the predictions of the models.

The extremes in the results of the comparison are represented by LFSM and GMOSS, which in general predict the brightest and faintest skies, respectively. In the frequency range from 30 to \SI{80}{MHz}, ULSA produces the second brightest sky maps. Interestingly, this is the frequency range, in which the set of LWA1 reference maps lie that are used in both LFSM and ULSA but no other model. In light of this, it would be interesting to investigate whether the calibration of the LWA1 has a systematic bias towards higher temperatures. In general, however, the complex character of how the models are correlated with each other in multiple ways makes it very hard to determine why predictions deviate and which models may be more trustworthy than others.

\begin{table*}
\caption{Value of $r_{\text{m}_1,\text{m}_2}$ listed for each combination of the interpolation models.}
\label{tab:model_comparison}
\centering
\begin{tabular}{c c c c c c c c}
\hline\hline
$r_{\text{m}_1,\text{m}_2}$ (\%) & LFmap & GSM & GSM16 & LFSM & GMOSS & SSM & ULSA \\ \hline
LFmap   &   -    	  &  $\SI{4.3}{ }$    &   $\SI{0.7}{ }$     &   $\SI{-7.3}{ }$  &   $\SI{7.0}{ }$   &   $\SI{-1.8}{ }$   &   $\SI{-4.8}{ }$ \\ 
GSM     &   $\SI{-4.3}{ }$    &   -  	  &   $\SI{-3.6}{ }$    &   $\SI{-11.7}{ }$   &   $\SI{2.7}{ }$   &   $\SI{-6.1}{ }$   &   $\SI{-9.1}{ }$ \\ 
GSM16   &   $\SI{-0.7}{ }$    &  $\SI{3.6}{ }$    &    -   		&   $\SI{-8.1}{ }$   &   $\SI{6.3}{ }$   &   $\SI{-2.5}{ }$   &   $\SI{-5.5}{ }$ \\ 
LFSM    &   $\SI{7.3}{ }$     &  $\SI{11.7}{ }$   &   $\SI{8.1}{ }$    &   -   &   $\SI{14.3}{ }$   &   $\SI{5.5}{ }$   &   $\SI{2.6}{ }$ \\
GMOSS    &   $\SI{-7.0}{ }$     &  $\SI{-2.7}{ }$   &   $\SI{-6.3}{ }$   &   $\SI{-14.3}{ }$   &   -    &   $\SI{-8.8}{ }$   &   $\SI{-11.8}{ }$ \\
SSM    &   $\SI{1.8}{ }$     &  $\SI{6.1}{ }$   &   $\SI{2.5}{ }$   &   $\SI{-5.5}{ }$   &   $\SI{8.8}{ }$   &   -    &   $\SI{-2.9}{ }$ \\
ULSA    &   $\SI{4.8}{ }$     &  $\SI{9.1}{ }$   &   $\SI{5.5}{ }$   &   $\SI{-2.6}{ }$   &   $\SI{11.8}{ }$   &   $\SI{2.9}{ }$   &   -  \\
\hline
\end{tabular}
\tablefoot{The header row gives the model m\textsubscript{1} and the leftmost column gives the model m\textsubscript{2}. Positive values show that the model m\textsubscript{1} produces maps with higher average temperatures than the model m\textsubscript{2}.}
\end{table*}

\subsection{Comparison of the local sky}
The differences between the models are not just reflected in general deviations of the temperature scales. They also show structural variations on larger scales that become noticeable by comparing specific sky regions, for instance, on/off the Galactic plane. These variations influence the model comparison when confining their output maps to the sky coverage of a specific radio-detection experiment on Earth. The local sky of an observer changes with the local sidereal time (LST) and solely depends on the observer's geographic latitude if the experiment is operated both day and night.

The average temperature of a local sky was obtained by converting a map into the horizontal coordinate system with the two angles azimuth $\alpha$ and altitude a (the elevation angle above horizontal), thus limiting it to the visible half of the sky above the horizon and integrating over this region. The results were then averaged for varying LST from \SI{0}{h} to \SI{24}{h} as

\begin{equation}
\label{eq:local_sky_t}
\begin{split}
& \bar{T}_\text{local}(\nu, \ell) = \\
& \frac{1}{2\pi} \frac{1}{\SI{24}{h}} \int_{\SI{0}{h}}^{\SI{24}{h}} \text{d} t_{\text{LST}} \int_{0}^{\pi} \text{d} a \int_{\frac{-\pi}{2}}^{\frac{\pi}{2}} \text{d} \alpha \cos{(a)}\; T(\nu, \ell, t_{\text{LST}}; a, \alpha)
\end{split}
\end{equation}

for a given latitude $\ell$ of the observer on Earth. We further integrated the average temperature of the local sky over a frequency range $[\nu_\text{lower},\nu_\text{upper}]$:

\begin{equation}
\label{eq:local_sky_sigma_t}
\mathcal{T} (\ell) = \int_{\nu_\text{lower}}^{\nu_\text{upper}} \bar{T}_\text{local}(\nu, \ell) \; \text{d} \nu .
\end{equation}

\begin{figure*}
    \centering
    \includegraphics[width=17cm]{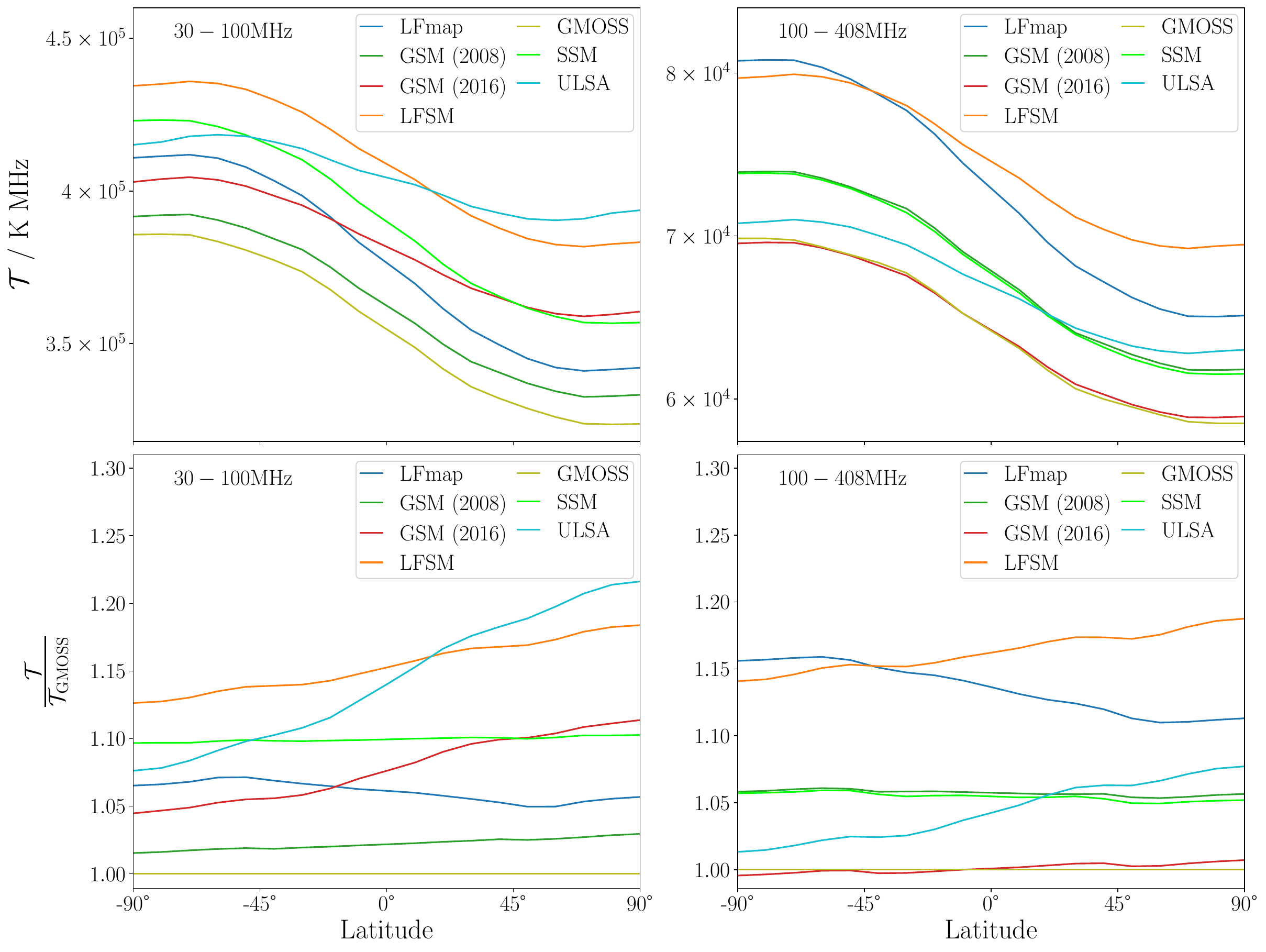}
    \caption{Values of $\mathcal{T} (\ell)$ for each sky interpolation models as a function of latitude. The results for the frequency ranges $[\SI{30}{MHz},\SI{100}{MHz}]$ and $[\SI{100}{MHz},\SI{408}{MHz}]$ are shown on the left and right side, respectively. In the top row, absolute values are shown while in the bottom row, values are normalized to GMOSS.}
    \label{fig:All_models_local_lat_combined_all}
\end{figure*}

The result of this integral is plotted for each interpolation model and for the two frequency ranges $[\SI{30}{MHz},\SI{100}{MHz}]$ and $[\SI{100}{MHz},\SI{408}{MHz}]$ as a function of the observer's latitude in Fig.\ \ref{fig:All_models_local_lat_combined_all} in the top row. It is clear that $\mathcal{T}$ changes depending on the exposure to the radio-bright center of the Galaxy. At the northern celestial pole ($\SI{90}{\degr}$ latitude) this exposure is the smallest and thus $\mathcal{T}$ is the lowest over all frequencies, while the maximum exposure takes place at around $\SI{-60}{\degr}$ of latitude.

Furthermore, differences between the interpolation models are recognizable. Apart from a general shift in the temperature scale, there are latitude dependent variations, which are probably due to different ratios in the models between the coldest and hottest regions of the sky.

The bottom row of Fig.\ \ref{fig:All_models_local_lat_combined_all} shows the results from the top row normalized to those of GMOSS, which on average gives the coldest sky maps. Here the overall level of agreement is found to be around 20\% for latitudes around the north celestial pole and around 13\% around the south celestial pole. From the variation of the level of agreement across the geographical latitude, we conclude that the reference maps of the models, which often do not cover the full range in latitude and therefore affect the predictions in only limited parts of the sky, have a major role in the total uncertainty of how well the diffuse Galactic emission can be predicted.

Comparing the results for the two chosen frequency ranges, the models rank differently in the sky brightness they predict. However, the relative spread between faintest and brightest model remains the same, as can be seen from the plots with normalization. 

\section{Comparison for selected radio experiments}
The comparison of the sky models based on the local sky was further carried out for selected radio arrays for the detection of cosmic particles, namely \mbox{RNO-G} (surface antennas) \citep{RNO-G_Design}, LOFAR \citep{Schellart_2013}, GRAND \citep{GRAND_Design}, \mbox{OVRO-LWA} \citep{Monroe_2020}, \mbox{SKA-low} \citep{Buitink_2021}, the Pierre Auger Observatory (AERA \citep{AERA} and the \mbox{AugerPrime} Radio Detector \citep{Pont_2019}), and the radio antennas of the \mbox{IceCube} surface array \citep{IceCube_SurfaceArray_Development}. The sky models compare differently for each of the experiments. This is due to them being situated at different geographical latitudes and using different frequency bands. For this study, we made the simplifying assumption of a constant antenna sensitivity over the whole frequency band for all arrays. Further, we also restricted the local sky seen by the antennas to altitude angles between \SI{15}{\degr} and \SI{90}{\degr} above the horizon to simulate a uniform gain pattern in that range. This gain pattern is a toy model, that attempts to resemble the typically wide beam widths of radio antennas used in astroparticle physics. Using array-specific antenna patterns, obtained either from simulations or data, would be favorable. However, most of this information is not publicly available, if at all (e.g.,\ for arrays under construction or development). At this level of detail, one would need to take into account frequency-dependent responses of the electronic components in the signal chain as well, which is beyond the scope of this work and we leave that to dedicated analyses of the Galactic calibration of the individual arrays. Therefore, we also ignored a potential polarization of the diffuse radio emission and polarization characteristics of the individual antennas.

With the antenna toy model, the average temperature of the local sky integrated over LST is calculated analogous to Eq.\ \ref{eq:local_sky_t} as

\begin{equation}
\begin{split}
& \bar{T}_{\text{exp}}(\nu, \ell_\text{exp}) = \\
& \frac{1}{2\pi} \frac{1}{\SI{24}{h}} \int_{\SI{0}{h}}^{\SI{24}{h}} \mathrm{d} t_{\text{LST}} \int_{\SI{15}{\degree}}^{\SI{90}{\degree}} \cos{(a)} \mathrm{d} a \int_{\frac{-\pi}{2}}^{\frac{\pi}{2}} T(\nu, \ell_\text{exp}, t_{\text{LST}}; a, \alpha) \mathrm{d} \alpha ,
\end{split}
\end{equation}

with the latitude of the respective experiment $\ell_\text{exp}$. We integrated $\bar{T}_{\text{exp}}(\nu, \ell_\text{exp})$ over the corresponding frequency band [$\nu_\text{exp, lower}$, $\nu_\text{exp, upper}]$ to obtain

\begin{equation}
\mathcal{T_\text{exp}} (\ell_\text{exp}) = \int_{\nu_\text{exp, lower}}^{\nu_\text{exp, upper}} \bar{T}_{\text{exp}}(\nu, \ell_\text{exp}) \; \text{d} \nu .
\end{equation}

Further, any two sky models $\mathrm{m}_1$ and $\mathrm{m}_2$ were compared by their relative difference

\begin{equation}
r_{\text{exp; m}_1\text{, m}_2} = 2 \frac{ \mathcal{T}_{\text{exp, m}_1} (\ell_\text{exp}) - \mathcal{T}_{\text{exp, m}_2} (\ell_\text{exp}) }{ \mathcal{T}_{\text{exp, m}_1} (\ell_\text{exp}) + \mathcal{T}_{\text{exp, m}_2} (\ell_\text{exp}) }.
\end{equation}

The largest relative differences obtained for each experiment are listed in Table \ref{tab:experiments_comparison} together with the properties of the experiment and the combination of sky models between which the largest relative difference occurs. These differences range from $\sim12\%$ to $\sim22\%$. Again, the brightest and faintest skies are predicted mostly by LFSM and GMOSS, respectively. However, because of the unique locations and frequency bands, the models in the roles of the extrema are different for \mbox{RNO-G}, LOFAR, and \mbox{OVRO-LWA}. In dedicated analyses of the arrays regarding the Galactic calibration, instead of using all models, the effort to predict the Galactic emission can be reduced to just using these border models. An extended overview of the comparison results for each of the experiments including every combination of models is given in appendix \ref{chap:appendixA}.

\begin{table*}
    \caption{Largest relative deviation $\max(r_{\text{Exp; m}_1\text{, m}_2})$ between any two sky models m\textsubscript{1} and m\textsubscript{2} for each of the selected radio experiments.}
    \label{tab:experiments_comparison}
    \centering
    \begin{tabular}{c c c c c}
    \hline\hline
    Experiment / observatory & $\ell_\text{exp}$ (\degr) & Frequency band (MHz) & $\max(r_{\text{exp; m}_1\text{, m}_2})$ (\%) & Corresponding sky models \\ \hline
    RNO-G (1)   & 72.58  & 100 to 408 & 17.1   & LFSM / GSM16   \\ 
    LOFAR low (2)   & 52.91  & 30 to 80  & 19.8   & ULSA / GMOSS  \\
    LOFAR high (3)   & 52.91  & 110 to 190  & 18.4   & LFSM / GSM16  \\
    GRAND (4)   & 42.93  & 50 to 200  & 21.5   & LFSM / GMOSS  \\
    OVRO-LWA (5)   & 37.23 & 30 to 80   & 19.3   & ULSA / GMOSS  \\
    SKA-low (6) & -26.70 & 50 to 350  & 15.1   & LFSM / GMOSS  \\
    Auger (7)   & -35.21 & 30 to 80   & 11.7   & LFSM / GMOSS  \\
    IceCube (8) & -90.0  & 70 to 350 & 20.3   & LFSM / GMOSS  \\
    \hline
    \end{tabular} 
    \tablefoot{Site locations and frequency bands of the experiments are quoted from latest design plans. For RNO-G the LPDA surface antennas were considered, which are sensitive up to ${\sim}\SI{1}{GHz}$, but the comparison was done only up to \SI{408}{MHz} because of the limitations of the LFmap and LFSM sky models. The complete table can be found in appendix \ref{chap:appendixA}.}
    \tablebib{
    (1) \citet{RNO-G_Design}; (2) \citet{Schellart_2013}; (3) \citet{Nelles_2015}; (4) \citet{GRAND_Design}; (5) \citet{Monroe_2020}; (6) \citet{Buitink_2021}; (7) \citet{AugerPrime_Radio}; (8) \citet{IceCube_SurfaceArray_Development}.
    }
\end{table*}

\section{Influence of other natural sources of radio emission} 
\label{sec:5_AdditionalSources}
The Galaxy is not the only natural source of background radio emission, although it is dominant for remote locations in the considered frequency range. In the following, we discuss contributions to the radio background by the quiet Sun, the ionosphere, and Jupiter and how they may influence the Galactic calibration of radio antenna arrays.

\subsection{The quiet Sun} 
Another source of radio emission in the sky is the Sun. Here, we studied the influence of the quiet Sun on the average sky brightness, which describes the continuous thermal emission of solar radiation. There is also a concept of the active Sun, which includes enhanced emission during sunspot activity as well as emission in the context of solar flares. These contributions take place on limited timescales from seconds to hours and may be significantly brighter than the quiet sun. The slowly varying thermal emission from the sun reaches up to two times the maximum brightness temperature of the quiet state, whereas the brightness temperature of the rapidly varying solar emission components may be five or more magnitudes larger than the quiet state, depending on the emission mechanism \citep{Kraus_RadioAstronomy}. However, we did not cover the active sun in this study.

Radio emission of the quiet Sun at frequencies from tens of megahertz to tens of gigahertz is larger than expected from a \SI{6000}{K} black body spectrum and reaches brightness temperatures up to \SI{e6}{K} around \SI{50}{MHz} \citep{Kraus_RadioAstronomy}. To investigate how large the influence of the Sun is, we generated sky maps with a disk of constant brightness projected onto it. Data on the brightness temperature of the quiet Sun as well as its effective size as a function of the frequency were taken from a summary of recent measurements given in \citep{Zhang_2022}. In the considered frequency range, the quiet Sun appears in a rather elliptical shape and the data of the measured radius are given in the north-south and east-west direction separately. We determined a conservative estimate for the solar influence on the average sky temperature by modeling the Sun as a circular disk using the larger radius data of the east-west direction.

\begin{figure}
	\resizebox{\hsize}{!}{\includegraphics{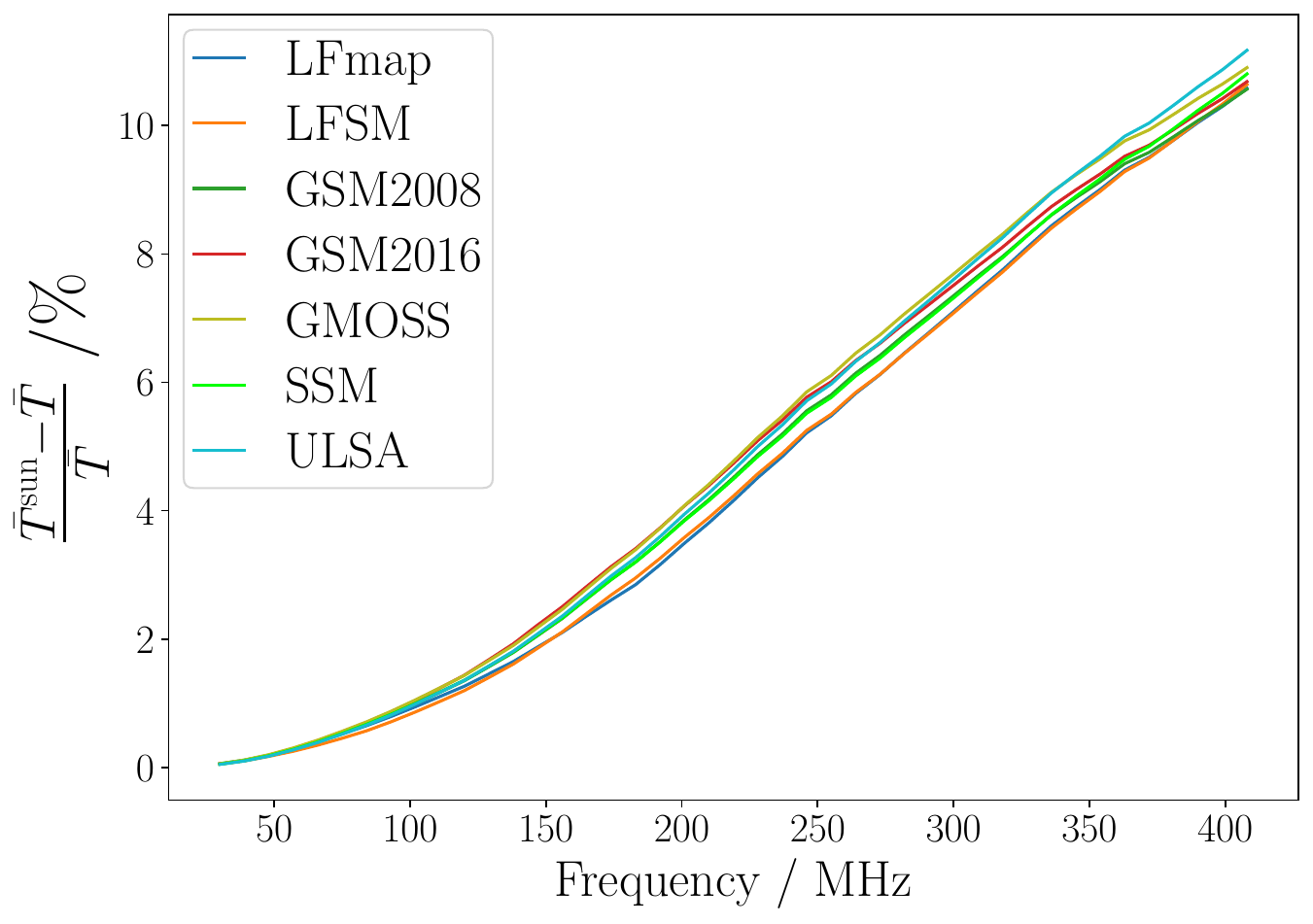}}
	\caption{Influence of the Sun on the average sky temperature for each of the models. The plot shows the relative difference of the average sky temperature as a function of the frequency for when a Sun sized circle of the corresponding brightness temperature (quiet Sun) is added to the maps. Kinks in the curves are due to quantization effects in the calculations.}
	\label{fig:All_models_sun}
\end{figure}

With the superimposed quiet Sun the average sky temperature of the map was calculated as before and its relative difference to the average sky temperature of the unmodified map was determined. This is shown for each of the sky interpolation models in Fig.\ \ref{fig:All_models_sun}. While the influence is negligible at the lowest frequencies, it grows to a level of $\sim11\%$ at \SI{400}{MHz}.

\begin{figure}
    \centering
	\resizebox{\hsize}{!}{\includegraphics{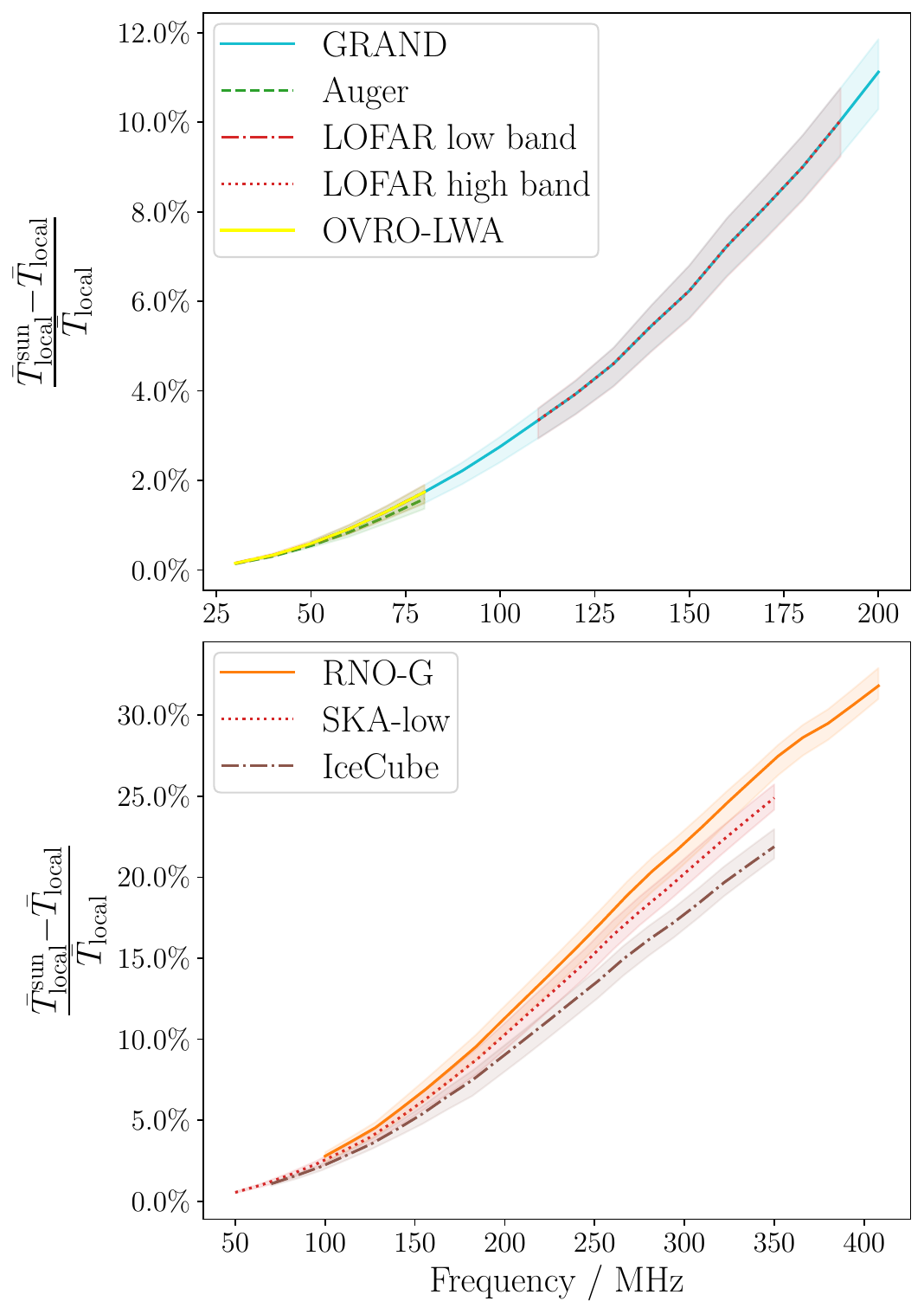}}
	\caption{Relative difference of the average temperature of the local sky induced by the quiet Sun for the selected radio arrays. The lines represent the average results from using the seven sky models to produce the maps, while the maximum and minimum contribution from any model is shown by the colored bands. Arrays at lower frequencies are shown in the top part, while arrays at higher frequencies are shown in the bottom part. Kinks in the curves are due to quantization effects in the calculations.}
	\label{fig:All_models_local_sun}
\end{figure}

Furthermore, we adapted this procedure to the local skies of the selected radio arrays, again restricting them to elevations from \SI{15}{\degr} to \SI{90}{\degr} above the horizon and taking the mean of the average sky temperatures over the course of \SI{24}{h} of LST. The relative difference caused by the Sun is shown for all arrays over their frequency bands in Fig.\ \ref{fig:All_models_local_sun}. The arrays at the lowest frequencies are shown on the left and the ones reaching to larger frequencies are shown on the right.

Analogous to Fig.\ \ref{fig:All_models_sun}, the relative differences increase with frequency, while they are scaled up here because only a portion of the total sky is considered. The differences are on the level of 1\% for Auger, \mbox{OVRO-LWA}, and the low band of LOFAR and go up to 30\% for \mbox{RNO-G} at around \SI{400}{MHz}. However, we note that also for the arrays at lowest frequencies the quiet Sun might become relevant, if the respective antenna's gain pattern is narrower than in our toy model, increasing the relative solar contribution during the time of passage through the main lobe of the beam pattern. Deviations between the arrays at the same frequencies are attributable to the different exposure to the Galactic center due to their geographical positions. 

\subsection{The ionosphere} \label{sec:5_Ionosphere}
The ionosphere is a reflective medium for radio waves below \SI{20}{MHz} \citep{Allan_1971}. Therefore, at these frequencies, noise from sources in the atmosphere (for example from thunderstorms) can be picked up from large distances by radio antennas, and would in fact dominate over the Galactic signal. The ionosphere's ion density and consequently the atmospheric noise is expected to be the largest during summer daytime in phases with high sunspot activities \citep{Allan_1971, ITU_Noise}.

Radio antenna arrays in astroparticle physics are typically operated at frequencies above \SI{30}{MHz} to avoid the atmospheric noise due to reflections from the ionosphere \citep{Huege_2016}. However, even at frequencies above \SI{30}{MHz}, the background signal measured in a radio detector may be affected by refractive and absorptive effects of the ionosphere. It is reported that there are daily variations of the ionosphere's opacity for radio waves \citep{Rogers_2015} but they can also appear on more irregular timescales, such as in the sporadic E-layer \citep{Davies_1965}. These variations may have an impact on the sky-averaged measurements in the Galactic calibration. While those ionospheric effects are significant for the detection of EoR signals \citep{Datta_2016}, detailed studies are necessary to determine whether they are relevant for the calibration of radio detectors in astroparticle physics. A useful web-tool by \citet{Rodland_2022}, called prop, shows real-time maps of the critical frequency up to which radio waves get bounced off of the ionosphere. This tool could help determining whether there is ionospheric influence on the measurements of a radio detector at a given time.

\subsection{Jupiter}
Jupiter is a source of radio emission with a unique and dynamic spectrum dominated by several components of nonthermal emission. This spectrum has a hard cutoff at \SI{40}{MHz} up to which frequency Jupiter emits intense decametric (DAM) radio waves \citep{Zarka_2005}. Jovian DAM emission occurs in irregular intervals with time scales of minutes and hours \citep{Cecconi_2012}. The antennas of the radio detectors considered in this work have a rather wide beam and a potential signal from Jupiter would always enter these together with a large-scale contribution from the Galaxy. Because of this and due the cutoff in its emission spectrum, Jupiter can be safely ignored above \SI{40}{MHz}. Below \SI{40}{MHz} we cannot give a definitive answer as to how Jovian emission influences the Galactic calibration. We advocate a dedicated analysis that takes into account the antenna gain pattern as well as the temporal features of Jupiter's spectrum. The probability for an observer to see Jovian emission at a specific time can be predicted well, for example using the Jupiter Radio Probability Tool \citep{Cecconi_2023}. We leave this to future work.

\section{Discussion} 
\label{sec:6_Discussion}
This study gives estimates for the systematic uncertainty in predicting the diffuse Galactic radio emission in the \num{30} to \SI{408}{MHz} range. Moreover, it provides an overview of the available models that are used for making these predictions and their corresponding reference measurements. A considerable realization from this overview is that probably a significant level of correlation exists between the systematic uncertainties of the sky models and the reference maps. Some reference maps served as calibrators for other maps and have thus an increased weight in the sky interpolation models. In general, for the sky models considered here, there was no substantial attention paid to weighting the individual reference maps during the modeling process, which can lead to biases on the absolute temperature scale. These biases are almost impossible to untangle.

The generally small number of reference maps in the studied frequency range further complicates the situation. While many surveys included a careful treatment of their uncertainties, others lack this kind of thoroughness, which in some cases may just be a consequence of the age of these surveys and the unavailability of some tools and techniques at the time of their execution.

In the future, these difficulties can be coped with by including more and newer reference maps into the sky interpolation models and weighting them based on their inherited uncertainties. New sky surveys at megahertz frequencies are on the horizon or have been published in the recent past, including direct comparisons to other maps or sky models \citep{Singal_2023, Eastwood_2018, Kriele_2022}.

Regarding the application of the Galactic calibration for radio arrays in astroparticle physics, we argue that the viability of this method is given if systematic uncertainties are carefully taken care of, which means that the parameters of the considered radio array are important. From our studies, we see that the Galactic calibration exhibits larger uncertainties at higher frequencies. There, the Galaxy's contribution to the radio background diminishes and sources such as the Sun become stronger. In the case of an array with a large frequency band, it might be a solution to conduct the Galactic calibration only with night-time measurements to minimize systematic uncertainties. Solar emission on short timescales, which is part of the so-called active sun, can be much brighter than the quiet sun. Therefore, time periods with contributions from the active sun may have to be excluded from Galactic calibration campaigns to prevent distortions in the calibration.

\section{Conclusion} 
\label{sec:7_Conclusion}
We compared seven interpolation models that produce full-sky maps of the Galactic radio emission at frequencies between 30 and \SI{408}{MHz}. The models partially rely on the same reference measurements and differ to a greater or lesser extent by their interpolation approach. A summary of the used reference maps shows relative uncertainties on the temperature scales of up to 20\% and zero-level errors that can be as large as up to 11\% of the average temperature of the sky at that frequency. In a global comparison, we find that the predictions of the interpolation models all agree within 14.3\%, which we suggest as an estimate for the systematic uncertainty of the model predictions. This estimate incorporates both the individual temperature scale uncertainties of the reference maps and uncertainties because of different modeling methods. 

Furthermore, we compared the models based on the local sky at a given experiment's latitude on Earth. This step reveals further differences between the models and in combination with specific frequency bands of selected radio-detection arrays deviations vary considerably. Depending on the experiment, the level of agreement lies between 11.7\% and 21.5\%.

Additionally, we studied the influence of the quiet Sun on the radio background of the selected arrays and find it to be negligible at lower frequencies. However, the contribution increases to $\sim30\%$ at around \SI{400}{MHz}. While the strongest ionospheric contributions happen below \SI{30}{MHz}, which is the lowest frequency typically used by radio arrays, there are effects that can disturb measurements even above this frequency. Therefore, a dedicated study of those effects is needed. We do not expect emission from Jupiter to play a role in the applicability of the Galactic calibration above \SI{40}{MHz}. For frequencies below \SI{40}{MHz}, we suggest the conduction of a quantitative analysis about Jupiter's influence.

The relative uncertainties in sky temperature quoted above determine the signal power measured in radio detection arrays for cosmic particles -- whereas the energy scale for the energy of the detected particles scales with the electric field amplitudes, that is, the square root of the signal power. The relevant systematic uncertainty on the energy scale of particle detection thus corresponds to approximately half of the values quoted here. These uncertainties are typically competitive with those achieved with external calibration sources -- if not better -- and thus confirm the value of the Galactic calibration approach. If models with even higher accuracy for the Galactic emission in the frequency band from 30 to \SI{408}{MHz} become available in the future, radio detection arrays for cosmic particles will profit from these retroactively.

\begin{acknowledgements}
We wish to thank the authors of the models for making their work public and helping with questions about details of the models. We also wish to thank Danny Price for providing the PyGDSM package, that made assessing the models a lot easier. Further, we wish to thank the referee of this article because their constructive feedback with their point of view from the radio astronomy community considerably improved the quality of this study. This research was supported by the Bundesministerium für Bildung und Forschung (BMBF) under the contract 05A20VK1.
\end{acknowledgements}

\bibliographystyle{aa}
\bibliography{References}

\begin{appendix} 
\section{Complete model comparison for selected radio experiments}
\label{chap:appendixA}

\begin{table*}[b]
\label{tab:extended_ratios_experiments}
\caption{Values of $r_{\text{exp; m}_1\text{, m}_2}$ for each combination of the interpolation models and for all selected radio arrays.}
\centering
\begin{tabular}{c c c c c c c c}
\hline\hline
\multicolumn{8}{c}{RNO-G}\\
   $r_{\text{exp; m}_1\text{, m}_2}$ (\%)     & LFmap & GSM & GSM16 & LFSM & GMOSS & SSM & ULSA \\ \hline
LFmap   &   -    	    &  $\SI{5.1}{ }$ &   $\SI{9.6}{ }$ &   $\SI{-7.6}{ }$  & $\SI{9.2}{ }$  & $\SI{5.8}{ }$  & $\SI{2.2}{ }$ \\
GSM     &   $\SI{-5.1}{ }$ &   -  	       &   $\SI{4.5}{ }$ &   $\SI{-12.7}{ }$  & $\SI{4.1}{ }$  & $\SI{0.7}{ }$  & $\SI{-2.9}{ }$ \\
GSM16   &   $\SI{-9.6}{ }$ &  $\SI{-4.5}{ }$ &    -   	   &   $\SI{-17.1}{ }$  & $\SI{-0.4}{ }$  & $\SI{-3.7}{ }$  & $\SI{-7.4}{ }$ \\
LFSM    &   $\SI{7.6}{ }$ &  $\SI{12.7}{ }$ &   $\SI{17.1}{ }$ &   -   & $\SI{16.8}{ }$  & $\SI{13.4}{ }$  & $\SI{9.8}{ }$  \\
GMOSS    &   $\SI{-9.2}{ }$ &  $\SI{-4.1}{ }$ &   $\SI{0.4}{ }$  & $\SI{-16.8}{ }$  &  -  & $\SI{-3.4}{ }$  & $\SI{-7.0}{ }$  \\ 
SSM      &   $\SI{-5.8}{ }$ &  $\SI{-0.7}{ }$ &   $\SI{3.7}{ }$  & $\SI{-13.4}{ }$  & $\SI{3.4}{ }$  &  -  & $\SI{-3.6}{ }$  \\ 
ULSA     &   $\SI{-2.2}{ }$ &  $\SI{2.9}{ }$ &   $\SI{7.4}{ }$  & $\SI{-9.8}{ }$  & $\SI{7.0}{ }$  & $\SI{3.6}{ }$  &  -  \\ \hline

\multicolumn{8}{c}{LOFAR low}\\ 
   $r_{\text{exp; m}_1\text{, m}_2}$ (\%)     & LFmap & GSM & GSM16 & LFSM & GMOSS & SSM & ULSA  \\ \hline
LFmap   &   -    	    &  $\SI{3.4}{ }$ &   $\SI{-5.5}{ }$ &   $\SI{-9.8}{ }$  & $\SI{6.1}{ }$  & $\SI{-3.7}{ }$  & $\SI{-13.8}{ }$ \\
GSM     &   $\SI{-3.4}{ }$ &   -  	       &   $\SI{-8.9}{ }$ &   $\SI{-13.2}{ }$  & $\SI{2.6}{ }$  & $\SI{-7.1}{ }$  & $\SI{-17.2}{ }$ \\
GSM16   &   $\SI{5.5}{ }$ &  $\SI{8.9}{ }$ &    -   	   &   $\SI{-4.3}{ }$  & $\SI{11.5}{ }$  & $\SI{1.8}{ }$  & $\SI{-8.3}{ }$ \\
LFSM    &   $\SI{9.8}{ }$ &  $\SI{13.2}{ }$ &   $\SI{4.3}{ }$ &   -    & $\SI{15.8}{ }$  & $\SI{6.1}{ }$  & $\SI{-4.0}{ }$ \\ 
GMOSS    &   $\SI{-6.1}{ }$ &  $\SI{-2.6}{ }$ &   $\SI{-11.5}{ }$  & $\SI{-15.8}{ }$  &  -  & $\SI{-9.7}{ }$  & $\SI{-19.8}{ }$  \\ 
SSM      &   $\SI{3.7}{ }$ &  $\SI{7.1}{ }$ &   $\SI{-1.8}{ }$  & $\SI{-6.1}{ }$  & $\SI{9.7}{ }$  &  -  & $\SI{-10.1}{ }$  \\ 
ULSA     &   $\SI{13.8}{ }$ &  $\SI{17.2}{ }$ &   $\SI{8.3}{ }$  & $\SI{4.0}{ }$  & $\SI{19.8}{ }$  & $\SI{10.1}{ }$  &  -  \\ \hline

\multicolumn{8}{c}{LOFAR high}\\
  $r_{\text{exp; m}_1\text{, m}_2}$ (\%)    & LFmap & GSM & GSM16 & LFSM & GMOSS & SSM & ULSA  \\ \hline
LFmap & -    	    &  $\SI{7.0}{ }$ &   $\SI{12.8}{ }$ &   $\SI{-5.6}{ }$  & $\SI{12.5}{ }$  & $\SI{8.1}{ }$  & $\SI{4.9}{ }$ \\
GSM &  $\SI{-7.0}{ }$ &   -  	       &   $\SI{5.9}{ }$ &   $\SI{-12.6}{ }$  & $\SI{5.6}{ }$  & $\SI{1.1}{ }$  & $\SI{-2.1}{ }$ \\
GSM16 &  $\SI{-12.8}{ }$ &  $\SI{-5.9}{ }$ &    -   	   &   $\SI{-18.4}{ }$  & $\SI{-0.3}{ }$  & $\SI{-4.8}{ }$  & $\SI{-8.0}{ }$ \\
LFSM &  $\SI{5.6}{ }$ &  $\SI{12.6}{ }$ &   $\SI{18.4}{ }$ &   -    & $\SI{18.1}{ }$  & $\SI{13.7}{ }$  & $\SI{10.5}{ }$ \\ 
GMOSS    &   $\SI{-12.5}{ }$ &  $\SI{-5.6}{ }$ &   $\SI{0.3}{ }$  & $\SI{-18.1}{ }$  &  -  & $\SI{-4.5}{ }$  & $\SI{-7.7}{ }$  \\ 
SSM      &   $\SI{-8.1}{ }$ &  $\SI{-1.1}{ }$ &   $\SI{4.8}{ }$  & $\SI{-13.7}{ }$  & $\SI{4.5}{ }$  &  -  & $\SI{-3.2}{ }$  \\ 
ULSA     &   $\SI{-4.9}{ }$ &  $\SI{2.1}{ }$ &   $\SI{8.0}{ }$  & $\SI{-10.5}{ }$  & $\SI{7.7}{ }$  & $\SI{3.2}{ }$  &  -  \\ \hline

\multicolumn{8}{c}{GRAND}\\
  $r_{\text{exp; m}_1\text{, m}_2}$ (\%)      & LFmap & GSM & GSM16 & LFSM & GMOSS & SSM & ULSA  \\ \hline
LFmap   &   -    	    &  $\SI{1.8}{ }$ &   $\SI{1.5}{ }$ &   $\SI{-14.3}{ }$  & $\SI{7.3}{ }$  & $\SI{1.4}{ }$  & $\SI{-5.2}{ }$ \\
GSM     &   $\SI{-1.8}{ }$ &   -  	       &   $\SI{-0.2}{ }$ &   $\SI{-16.0}{ }$  & $\SI{5.5}{ }$  & $\SI{-0.4}{ }$  & $\SI{-6.9}{ }$ \\
GSM16   &   $\SI{-1.5}{ }$ &  $\SI{0.2}{ }$ &    -   	   &   $\SI{-15.8}{ }$  & $\SI{5.7}{ }$  & $\SI{-0.1}{ }$  & $\SI{-6.7}{ }$ \\
LFSM    &   $\SI{14.3}{ }$ &  $\SI{16.0}{ }$ &   $\SI{15.8}{ }$ &   -     & $\SI{21.5}{ }$  & $\SI{15.7}{ }$  & $\SI{9.1}{ }$ \\
GMOSS    &   $\SI{-7.3}{ }$ &  $\SI{-5.5}{ }$ &   $\SI{-5.7}{ }$  & $\SI{-21.5}{ }$  &  -  & $\SI{-5.9}{ }$  & $\SI{-12.4}{ }$  \\ 
SSM      &   $\SI{-1.4}{ }$ &  $\SI{0.4}{ }$ &   $\SI{0.1}{ }$  & $\SI{-15.7}{ }$  & $\SI{5.9}{ }$  &  -  & $\SI{-6.6}{ }$  \\ 
ULSA     &   $\SI{5.2}{ }$ &  $\SI{6.9}{ }$ &   $\SI{6.7}{ }$  & $\SI{-9.1}{ }$  & $\SI{12.4}{ }$  & $\SI{6.6}{ }$  &  -  \\ \hline

\multicolumn{8}{c}{OVRO-LWA}\\
  $r_{\text{exp; m}_1\text{, m}_2}$ (\%)      & LFmap & GSM & GSM16 & LFSM & GMOSS & SSM & ULSA  \\ \hline
LFmap   &   -    	    &  $\SI{2.2}{ }$ &   $\SI{-5.8}{ }$ &   $\SI{-11.4}{ }$  & $\SI{4.6}{ }$  & $\SI{-5.2}{ }$  & $\SI{-14.8}{ }$ \\
GSM     &   $\SI{-2.2}{ }$ &   -  	       &   $\SI{-8.0}{ }$ &   $\SI{-13.6}{ }$  & $\SI{2.4}{ }$  & $\SI{-7.4}{ }$  & $\SI{-17.0}{ }$ \\
GSM16   &   $\SI{5.8}{ }$ &  $\SI{8.0}{ }$ &    -   	   &   $\SI{-5.6}{ }$  & $\SI{10.4}{ }$  & $\SI{0.6}{ }$  & $\SI{-9.0}{ }$ \\
LFSM    &   $\SI{11.4}{ }$ &  $\SI{13.6}{ }$ &   $\SI{5.6}{ }$ &   -     & $\SI{16.0}{ }$  & $\SI{6.2}{ }$  & $\SI{-3.4}{ }$ \\
GMOSS    &   $\SI{-4.6}{ }$ &  $\SI{-2.4}{ }$ &   $\SI{-10.4}{ }$  & $\SI{-16.0}{ }$  &  -  & $\SI{-9.8}{ }$  & $\SI{-19.3}{ }$  \\ 
SSM      &   $\SI{5.2}{ }$ &  $\SI{7.4}{ }$ &   $\SI{-0.6}{ }$  & $\SI{-6.2}{ }$  & $\SI{9.8}{ }$  &  -  & $\SI{-9.6}{ }$  \\ 
ULSA     &   $\SI{14.8}{ }$ &  $\SI{17.0}{ }$ &   $\SI{9.0}{ }$  & $\SI{3.4}{ }$  & $\SI{19.3}{ }$  & $\SI{9.6}{ }$  &  -  \\ \hline

\multicolumn{8}{c}{SKA-low}\\
   $r_{\text{exp; m}_1\text{, m}_2}$ (\%)   & LFmap & GSM & GSM16 & LFSM & GMOSS & SSM & ULSA  \\ \hline
LFmap & -    	    &  $\SI{4.8}{ }$ &   $\SI{7.5}{ }$ &   $\SI{-6.1}{ }$  & $\SI{9.1}{ }$  & $\SI{3.4}{ }$  & $\SI{6.1}{ }$ \\
GSM  &  $\SI{-4.8}{ }$ & -    	    &   $\SI{2.7}{ }$ &   $\SI{-10.9}{ }$  & $\SI{4.2}{ }$  & $\SI{-1.5}{ }$  & $\SI{1.2}{ }$ \\
GSM16 &  $\SI{-7.5}{ }$ &  $\SI{-2.7}{ }$ &    -   	   &   $\SI{-13.6}{ }$  & $\SI{1.6}{ }$  & $\SI{-4.1}{ }$  & $\SI{-1.5}{ }$ \\
LFSM &  $\SI{6.1}{ }$ &  $\SI{10.9}{ }$ &   $\SI{13.6}{ }$ &   -    & $\SI{15.1}{ }$  & $\SI{9.5}{ }$  & $\SI{12.2}{ }$ \\ 
GMOSS    &   $\SI{-9.1}{ }$ &  $\SI{-4.2}{ }$ &   $\SI{-1.6}{ }$  & $\SI{-15.1}{ }$  &  -  & $\SI{-5.7}{ }$  & $\SI{-3.0}{ }$  \\ 
SSM      &   $\SI{-3.4}{ }$ &  $\SI{1.5}{ }$ &   $\SI{4.1}{ }$  & $\SI{-9.5}{ }$  & $\SI{5.7}{ }$  &  -  & $\SI{2.7}{ }$  \\ 
ULSA     &   $\SI{-6.1}{ }$ &  $\SI{-1.2}{ }$ &   $\SI{1.5}{ }$  & $\SI{-12.2}{ }$  & $\SI{3.0}{ }$  & $\SI{-2.7}{ }$  &  -  \\ \hline

\end{tabular}
\end{table*}

\FloatBarrier

\begin{table*}
\ContinuedFloat
\caption{continued.}
\label{tab:extended_ratios_experiments_2}
\centering
\begin{tabular}{c c c c c c c c}
\hline\hline
\multicolumn{8}{c}{Auger}\\
   $r_{\text{exp; m}_1\text{, m}_2}$ (\%)     & LFmap & GSM & GSM16 & LFSM & GMOSS & SSM & ULSA  \\ \hline
LFmap   &   -    	    &  $\SI{4.8}{ }$ &   $\SI{0.7}{ }$ &   $\SI{-5.6}{ }$  & $\SI{6.0}{ }$  & $\SI{-3.1}{ }$  & $\SI{-2.1}{ }$ \\ 
GSM     &   $\SI{-4.8}{ }$ &   -  	    &   $\SI{-4.1}{ }$ &   $\SI{-10.4}{ }$  & $\SI{1.3}{ }$  & $\SI{-7.9}{ }$  & $\SI{-6.9}{ }$ \\ 
GSM16   &   $\SI{-0.7}{ }$ &  $\SI{4.1}{ }$ &    -   	   &   $\SI{-6.3}{ }$  & $\SI{5.4}{ }$  & $\SI{-3.8}{ }$  & $\SI{-2.8}{ }$ \\
LFSM    &   $\SI{5.6}{ }$ &  $\SI{10.4}{ }$ &   $\SI{6.3}{ }$ &   -    & $\SI{11.7}{ }$  & $\SI{2.5}{ }$  & $\SI{3.5}{ }$  \\
GMOSS    &   $\SI{-6.0}{ }$ &  $\SI{-1.3}{ }$ &   $\SI{-5.4}{ }$  & $\SI{-11.7}{ }$  &  -  & $\SI{-9.2}{ }$  & $\SI{-8.1}{ }$  \\ 
SSM      &   $\SI{3.1}{ }$ &  $\SI{7.9}{ }$ &   $\SI{3.8}{ }$  & $\SI{-2.5}{ }$  & $\SI{9.2}{ }$  &  -  & $\SI{1.0}{ }$  \\ 
ULSA     &   $\SI{2.1}{ }$ &  $\SI{6.9}{ }$ &   $\SI{2.8}{ }$  & $\SI{-3.5}{ }$  & $\SI{8.1}{ }$  & $\SI{-1.0}{ }$  &  -  \\ \hline

\multicolumn{8}{c}{IceCube}\\
   $r_{\text{exp; m}_1\text{, m}_2}$ (\%)   & LFmap & GSM & GSM16 & LFSM & GMOSS & SSM & ULSA  \\ \hline
LFmap & -    	    &  $\SI{9.0}{ }$ &   $\SI{13.6}{ }$ &   $\SI{-5.0}{ }$  & $\SI{15.3}{ }$  & $\SI{8.6}{ }$  & $\SI{11.0}{ }$ \\
GSM &  $\SI{-9.0}{ }$ &   -  	       &   $\SI{4.7}{ }$ &   $\SI{-13.9}{ }$  & $\SI{6.4}{ }$  & $\SI{-0.4}{ }$  & $\SI{2.0}{ }$ \\
GSM16 &  $\SI{-13.6}{ }$ &  $\SI{-4.7}{ }$ &    -   	   &   $\SI{-18.6}{ }$  & $\SI{1.7}{ }$  & $\SI{-5.0}{ }$  & $\SI{-2.7}{ }$ \\
LFSM &  $\SI{5.0}{ }$ &  $\SI{13.9}{ }$ &   $\SI{18.6}{ }$ &   -    & $\SI{20.3}{ }$  & $\SI{13.5}{ }$  & $\SI{15.9}{ }$ \\ 
GMOSS    &   $\SI{-15.3}{ }$ &  $\SI{-6.4}{ }$ &   $\SI{-1.7}{ }$  & $\SI{-20.3}{ }$  &  -  & $\SI{-6.8}{ }$  & $\SI{-4.4}{ }$  \\ 
SSM      &   $\SI{-8.6}{ }$ &  $\SI{0.4}{ }$ &   $\SI{5.0}{ }$  & $\SI{-13.5}{ }$  & $\SI{6.8}{ }$  &  -  & $\SI{2.4}{ }$  \\ 
ULSA     &   $\SI{-11.0}{ }$ &  $\SI{-2.0}{ }$ &   $\SI{2.7}{ }$  & $\SI{-15.9}{ }$  & $\SI{4.4}{ }$  & $\SI{-2.4}{ }$  &  -  \\ \hline

\end{tabular}
\end{table*}

\end{appendix}

\end{document}